\documentclass[fleqn,usenatbib]{mnras}
\pdfoutput=1
\usepackage{newtxtext,newtxmath}
\usepackage[T1]{fontenc}
\usepackage{ae,aecompl}

\usepackage{graphicx}	% Including figure files
\usepackage{amsmath}	% Advanced maths commands
\usepackage{multirow}
\usepackage{booktabs}
\usepackage{soul}
\usepackage{hyperref}
\usepackage[group-separator={,}]{siunitx}
\usepackage{ulem}

\graphicspath{{./figs/}}
\newcommand{\pc}{\,\mathrm{pc}}
\newcommand{\kpc}{\mbox{$\,\mathrm{kpc}$}}
\newcommand{\Msun}{\mbox{$\,\mathrm{M_{\odot}}$}}
\newcommand{\Gyr}{\mbox{$\,\mathrm{Gyr}$}}
\newcommand{\Myr}{\mbox{$\,\mathrm{Myr}$}}
\newcommand{\rper}{\mbox{$r_\mathrm{per}$}}
\newcommand{\rapo}{\mbox{$r_\mathrm{apo}$}}
\newcommand{\zmax}{\mbox{$z_\mathrm{max}$}}

\title[Co-formation of thin and thick discs]{Co-formation of the thin and thick discs revealed by  APOGEE-DR16 and {\it Gaia}-DR2}

\author[L. {Beraldo e Silva} et al.]{
Leandro {Beraldo e Silva}$^{1}$\thanks{E-mail: lberaldoesilva@gmail.com (LBS)},
Victor P. Debattista$^{1}$, 
David Nidever$^{2,3}$, \newauthor
Jo\~ao A. S. Amarante,$^{4,5}$ and
Bethany Garver$^{2}$\\
\\
$^{1}$Jeremiah Horrocks Institute, University of Central Lancashire,
Preston, PR1 2HE, UK\\
$^{2}$Department of Physics, Montana State University, P.O. Box 173840, Bozeman, MT 59717-3840\\
$^{3}$NSF's National Optical-Infrared Astronomy Research Laboratory, 950 North Cherry Ave, Tucson, AZ 85719 \\
$^{4}$Key Laboratory for Research in Galaxies and Cosmology, Shanghai  Astronomical Observatory,\\ Chinese Academy of Sciences, 80 Nandan Road, Shanghai 200030, China \\
$^{5}$ University of Chinese Academy of Sciences, No.19A Yuquan Road, Beijing 100049, China
}

\date{Accepted XXX. Received
  YYY; in original form ZZZ}

\pubyear{2020}

\begin{document}
\label{firstpage}
\pagerange{\pageref{firstpage}--\pageref{lastpage}}
\maketitle

% Abstract of the paper
\begin{abstract}
  Since thin disc stars are younger than thick disc stars on average,
  the thin disc is predicted by some models to start forming after the
  thick disc had formed, around 10 Gyr ago. Accordingly, no
  significant old thin disc population should exist. Using 6-D
  coordinates from Gaia-DR2 and age estimates from Sanders \& Das
  (2018), we select $\sim 24000$ old stars (${\tau > 10\Gyr}$, with
  uncertainties $\lesssim 15\%$) within $2\kpc$ from the Sun (full
  sample). A cross-match with APOGEE-DR16 ($\sim 1000$ stars) reveals
  comparable fractions of old chemically defined thin/thick disc
  stars. We show that the full sample pericenter radius (\rper)
  distribution has three peaks, one associated with the stellar halo
  and the other two having contributions from the thin/thick
  discs. Using a high-resolution $N$-body+SPH simulation, we
  demonstrate that one peak, at $\rper\approx 7.1\kpc$, is produced by
  stars from both discs which were born in the inner Galaxy and
  migrated to the Solar Neighbourhood. In the Solar Neighbourhood,
  $\sim 1/2$ ($\sim 1/3$) of the old thin (thick) disc stars are
  classified as migrators. Our results suggest that thin/thick discs
  are affected differently by radial migration inasmuch as they have
  different eccentricity distributions, regardless of vertical scale
  heights. We interpret the existence of a significant old thin disc
  population as evidence for an early co-formation of thin/thick
  discs, arguing that clump instabilities in the early disc offer a
  compelling explanation for the observed trends.
\end{abstract}

% Select between one and six entries from the list of approved keywords.
% Don't make up new ones.

%%%%%%%%%%%%%%%%%%%%%%%%%%%%%%%%%%%%%%%%%%%%%%%%%%

%%%%%%%%%%%%%%%%% BODY OF PAPER %%%%%%%%%%%%%%%%%%

\section{Introduction}
\label{sec:intro}

The Milky Way (MW) disc is a compound structure, as first revealed by
its double exponential vertical density profile
\citep[][]{1982PASJ...34..365Y, 1983MNRAS.202.1025G}. The thin disc is
characterized by stars on low-eccentricity orbits and which, on
average, are younger than the thick disc. In turn, the thick disc is
more centrally concentrated, has a larger fraction of stars on
eccentric orbits and is older on average \citep[see
e.g.][]{2012ApJ...753..148B,2019MNRAS.482.3426M,
  2020arXiv200303316C}. Additionally, the MW chemical abundance map
([$\alpha$/Fe] vs [Fe/H]) is bi-modal, with $\alpha$-rich stars being
more vertically extended, in contrast to $\alpha$-poor stars, and thus
associated with the geometrical thick disc
\citep[e.g.][]{2011A&A...535L..11A, 2014A&A...562A..71B,
  2014AA...564A.115A, 2014ApJ...796...38N}. However, the discs defined
chemically and geometrically are not identical; in this work we adopt
the chemical definition.

Because the thick disc is old, modelling its formation is
intrinsically related to dating the earliest events in the Galaxy's
evolution, and several possibilities have been considered, including
the early accretion from disrupted satellites
\citep[][]{2003ApJ...597...21A}, heating of a proto-disc by a major
merger \citep[][]{1993ApJ...403...74Q, 2008ApJ...688..254K,
  2008MNRAS.391.1806V, 2018Natur.563...85H}, in-situ star formation
following a gas-rich merger \citep[][]{2004ApJ...612..894B},
scattering by clumps in an early gas-rich phase
\citep[][]{2002MNRAS.330..707K, 2009ApJ...707L...1B,
  2019MNRAS.484.3476C, 10.1093/mnras/staa065}, radial migration
\citep[][]{2009MNRAS.399.1145S, 2011ApJ...737....8L}, direct in-situ
formation from a turbulent thick gas disc \citep[so called
``upside-down'' formation --][]{2013ApJ...773...43B}, mergers with
subsequent cosmological inflow of gas along filaments
\citep{2020arXiv200606008A, 2020arXiv200606011R, 2020arXiv200606012R}
and the rapidly changing orientation of the galactic plane
\citep{2020arXiv200610642M}.

While some of these scenarios (notably the upside-down) predict a
sequential formation, with the geometric thin disc significantly
forming only after the thick disc, in other scenarios the thin disc is
expected to start forming simultaneously with the thick disc, with the
consequence that a significant population of old thin disc stars
should exist. For instance, \cite{2003ApJ...597...21A} predict that
$\approx 15\%$ of currently (kinematically defined) thin disc stars
should have ages $\tau > 10$ Gyr in the accretion scenario, while in
the major merger picture, the geometric thin remnant of the heated
disc should represent $15\%-25\%$ of the total stellar mass at the end
of the merger, according to \cite{2008MNRAS.391.1806V}. Additionally,
in the clumpy formation scenario, the early disc develops clumps,
which heat stars vertically \protect\citep[as originally shown
by][]{2009ApJ...707L...1B} and enrich the medium with
$\alpha$-elements, producing a chemical bi-modality \citep[as shown
by][]{2019MNRAS.484.3476C} and geometric thin$+$thick discs similar to
those observed in the MW \citep[as shown
by][]{10.1093/mnras/staa065}. Thus, thin$+$thick disc co-formation and
an old thin disc population are natural predictions of this
scenario. Other models of thick disc formation may also suggest a
thin$+$thick co-formation \citep[see e.g.][]{2020MNRAS.497.2371L,
  2020arXiv200610195K, 2020arXiv200606008A, 2020arXiv200606011R}.

Whatever may be the case for the thin disc's early formation, there is
evidence for it growing inside-out at later times \citep[as in,
e.g.,][]{2001ApJ...554.1044C, 2007ApJ...658.1006M,
  2016ApJ...823...30B}, i.e. stars forming at progressively larger
radii, with the consequence that, at a fixed time, stars born at
smaller radii are typically metal-richer than those formed in the
disc's outer part. The observation of these trends is complicated by
radial migration (churning), which can move stars inwards or outwards
by several kpc on short timescales. This mechanism changes individual
stars angular momenta without changing their radial actions and is
more efficient for nearly-circular orbits
\citep[see][]{2002MNRAS.336..785S, 2012MNRAS.426.2089R}. An important
practical consequence, due to the density decreasing with radius, is a
net movement outward of more metal-rich inner stars, as detected,
e.g., in APOGEE \citep{2015ApJ...808..132H}, LAMOST+RAVE+{\it
  Gaia}-DR1 \citep{2018ApJ...860...91V} and in simulations
\citep{2016ApJ...818L...6L} -- see also \citet{2009MNRAS.396..203S}.

Recently, \cite{2020MNRAS.492.3408P} found evidence for a population
of 22 RR Lyrae stars \citep[which are older than 10 Gyr -- see
e.g.][]{2008AJ....135.1106G, 2013ApJ...775..134V} with pericenter
radii peaking at $\rper\approx 7$ kpc, small vertical excursions
$(\zmax < 0.9\kpc)$, high metallicities and low $\alpha$-abundances,
which they deemed a `conundrum': how can an old population have
chemical and kinematic properties typical of the thin disc? And how,
in apparent contradiction with an inside-out growth, would this old
population be located at such large radii?

In this work, using the catalogue of \cite{2018MNRAS.481.4093S}, in
conjunction with chemical abundances from APOGEE-DR16
\citep[][]{2020ApJS..249....3A, 2020arXiv200705537J}, we confirm the
existence of a significant population of old ($\tau > 10$ Gyr) thin
disc stars in the Solar Neighbourhood and the presence of a peak in
the pericenter radii at ${\rper\approx 7\, \mathrm{kpc}}$ with
contributions from both thin and thick discs. We further interpret the
existence of an old thin disc population as evidence for an early
co-formation of the thin and thick discs, with the $\rper$ peak
produced by radial migration of both components.  In
Sec. \ref{sec:obs_data}, we present our data sample and explain the
orbit integration procedure, as well as the assumed Galaxy model. In
Sec. \ref{sec:results} we present our results supporting an early
co-formation of the thin and thick discs. The location of the $\rper$
peak is investigated by means of a $N$-body+SPH simulation in
Sec. \ref{sec:sims}, where we show that these stars must have formed
in the inner disc and end up at larger radii due to radial
migration. In Sec.~\ref{sec:discuss} we discuss the implications of
this finding for different scenarios of the early evolution of the MW
and, and in particular, for the thick disc. We also discuss the
differences of radial migration in the thin and thick discs and how
our results solve the apparent conundrum of thin disc RR Lyrae
stars. Finally, we summarize our conclusions in
Sec.~\ref{sec:conclusions}.

%%%%%%%%%%%%%%%%%%%%%%%%%%%%%%%%%%%%%%%%%%%%%%%%%%

\section{Observational data and orbit integration}
\label{sec:obs_data}

We explore the astrometric data from {\it Gaia}-DR2 and age estimates
from the value-added catalogue of \cite{2018MNRAS.481.4093S}. This
catalogue is based on several spectroscopic surveys data and provides
distance, mass and age estimates for $\sim$ 3 million stars. It also
provides Galactocentric cylindrical coordinates and velocities and,
when available, IDs of stars identified in the surveys. We refer the
reader to the original paper for details.

We select stars in the subsamples of giants
(defined as ${\log g > 3\mathrm{dex}}$ and
${\log_{10}(T_\mathrm{eff}/K) < 3.73}$), and turnoff stars (defined as
${3.6 < \log g < 4.5}$ and ${\log_{10}(T_\mathrm{eff}/K) < 4.1}$), for
which ages are most reliable, according to
\cite{2018MNRAS.481.4093S}. We select old stars ($\tau >10$ Gyr), with
uncertainties ${\sigma_\tau < 1.5 \Gyr}$, within a distance
${d<2\kpc}$ from the Sun. Additionally, we require accurate parallaxes
(${|\varpi/\sigma_\varpi| > 5}$) and line-of-sight velocities
(${|v_\mathrm{los}/\sigma_{v_\mathrm{los}}| > 5}$). Note that this last
requirement suppresses stars with small $v_\mathrm{los}$, even if
$\sigma_{v_\mathrm{los}}$ is as small as 1 km/s, mainly suppressing
the contribution of near-by thin disc stars. Thus, this is a
conservative cut for the detection of a significant old thin disc
population and does not affect our conclusions. In order
to avoid stars with too small age uncertainties,
artificially produced by the model isochrone gridding, we require
$\sigma_{\log_{10}\tau} > 0.015$dex, as suggested by
\cite{2018MNRAS.481.4093S}. Finally, we restrict to stars with the
flag \texttt{best}=1 (stars with a match in {\it Gaia} and no
duplicate observations or observational problem detected). This
results in a sample of $\num{23857}$ stars, with a median ${\sigma_\tau \approx 0.99\Gyr}$ (and maximum ${\sigma_\tau = 1.5\Gyr}$, defined
in the quality cut).

We use the phase-space coordinates of our sample as initial conditions
and, with the help of the {\sc agama} package
\citep[][]{2019MNRAS.482.1525V}, integrate the orbits in the Galactic
potential of \cite{10.1093/mnras/stw2759}. We identify $62$ unbounded
stars, which we exclude, leaving a sample with $\num{23795}$ bound
stars. Each orbit is integrated forwards for 10 dynamical times
(i.e. for $\sim 2\Gyr$), and we calculate the pericenter and apocenter
radii, $\rper$ and $\rapo$, respectively. The eccentricity is
estimated as
\begin{equation}
  \label{eq:1}
  e = \frac{\rapo - \rper}{\rapo + \rper}.
\end{equation}
Finally, for each orbit we also calculate the maximum height from the
Galactic plane $\zmax$.

Additionally, when further splitting the data into thick and thin
discs, we cross-match with APOGEE-DR16 \citep[][]{2020ApJS..249....3A,
  2020arXiv200705537J}, resulting in a sample of $\sim \num{1000}$
stars with accurate metallicities and
$\alpha$-abundances. \cite{2019A&A...626A..16R} showed that magnesium
is the $\alpha$-element with most reliable abundance estimates in
APOGEE and we therefore use [Mg/Fe] to trace the $\alpha$-abundances.

% %%%%%%%%%%%%%%%%%%%%%%%%%%%%%%%%%%%%%%%%%%%%%%%%%%

\section{Observational results}
\label{sec:results}

We now show the results obtained from our sample of old stars in the
Solar Neighbourhood.  We start analyzing the distribution of
eccentricities and pericenter radii, demonstrating the presence of
three peaks, one of which is at $\rper\approx 7$ kpc. In
Sec.~\ref{sec:halo_thin_thick}, we cross-match the data with the
APOGEE-DR16, split the resulting sample into chemical thin and
thick discs and investigate the chemo-kinematic properties of these
components, showing their relative contributions to the $\rper$
peaks. In Sec.~\ref{sec:dissec_thin_thick} we present the first
evidence that the peak at $r_\mathrm{per}\approx 7$ kpc is produced by
radial migration. This is further investigated with a high-resolution
simulation in Sec.~\ref{sec:sims}.

%%%%%%%%%%%%%%%%%%%%%%%%%%%%%%%%%%%%%%%%%%%%%%%%%%
\subsection{Peaks in pericenter radius}
\label{sec:MW_peaks}

\begin{figure}
  \center
  \includegraphics[width=\columnwidth]{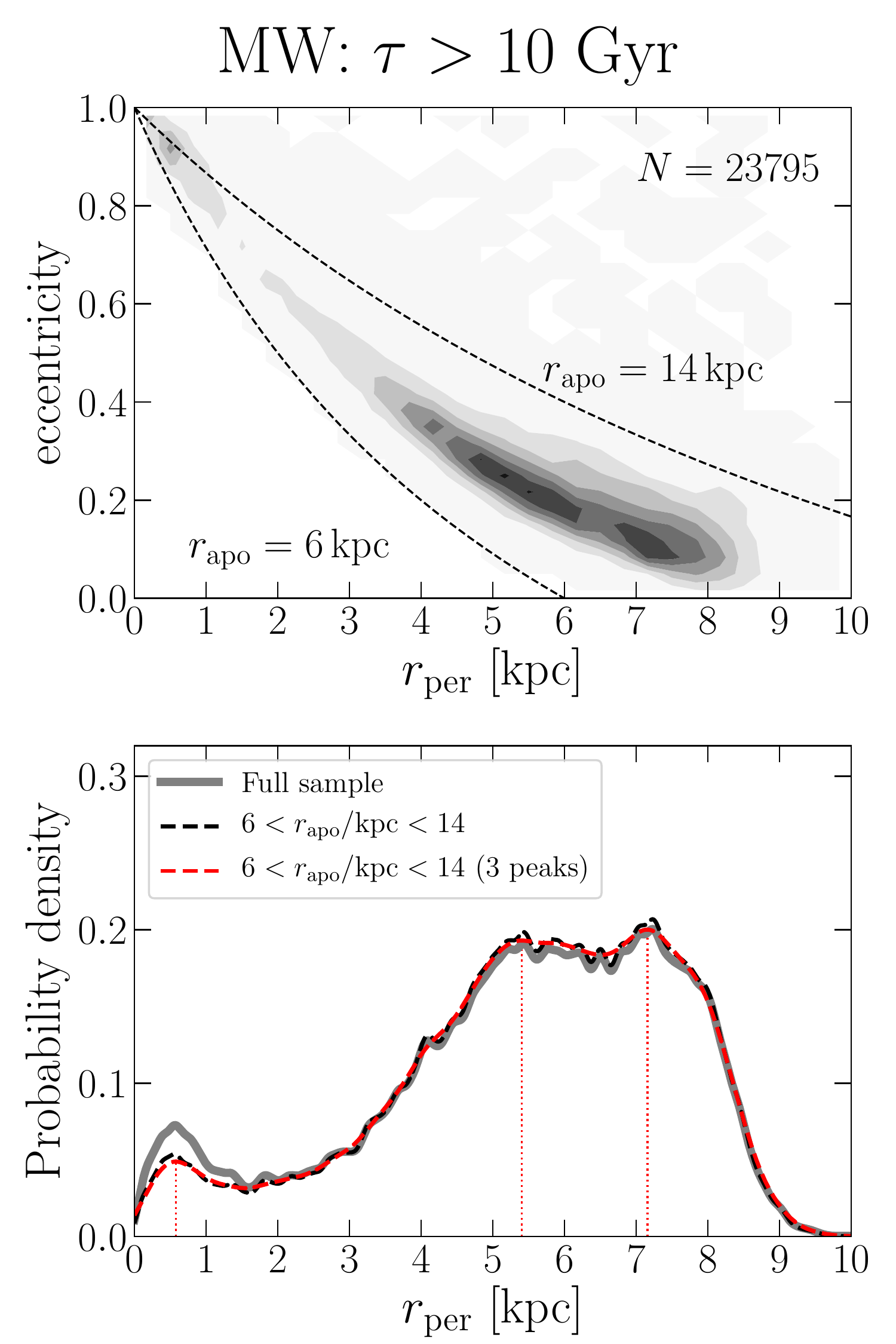}
  \caption{Top: distribution of old (turnoff + giant) stars in eccentricity vs pericentric radius, calculated with
    ages and astrometric data from
    \protect\cite{2018MNRAS.481.4093S} -- see
    Sec.~\ref{sec:obs_data}. Bottom: KDEs of $\rper$: for the full
    sample (grey), for $6 < \rapo/\mathrm{kpc} < 14$ (black dashed)
    and for this region and the most detailed KDE producing 3 peaks
    (dashed red) at the positions shown by vertical lines (see text for details). A statistical test rules out models with less
    than 3 peaks.}
  \label{fig:MW_e_vs_rper_all}
\end{figure}

Fig. \ref{fig:MW_e_vs_rper_all}, upper panel, shows contours for
eccentricity versus pericenter radius for our sample. The dashed lines
represent two fixed values of apocenter radius $\rapo$ -- see
Eq.~\ref{eq:1}. The absence of data with ${\rapo < 6\kpc}$ reflects
the selection of stars within ${d<2\kpc}$ from the Sun, given the
Solar position ${R_\odot \approx 8\kpc}$, while the vast majority of
stars have ${\rapo<14\kpc}$. The distribution peaks along a line of
approximately constant $\rapo$, in
${8 \lesssim \rapo/\kpc\lesssim 9}$, which also must be a selection
effect given the Solar position and the larger time stars spend near
their apocenters. The contours suggest the presence of three peaks:
the first at $(\rper/\kpc,e)\approx (0.5,0.9)$, the second at
$\approx (5-6, 0.25$) and the third at $\approx (7, 0.1)$.

The lower panel shows the kernel density estimate (KDE) of $\rper$
(i.e. of the projection of the upper panel on the $x$-axis) using a
Gaussian kernel with window width $h=0.07$, which was chosen by
cross-validation (solid grey), where the three peaks mentioned before
can be seen more clearly. In order to eliminate stars with extreme or
unrealistic kinematic properties, which can arise because of the
assumed model for the MW's potential, we select stars with
$6<\rapo/\mathrm{kpc}<14$, i.e. between the dashed curves in the upper
panel. The $\rper$ KDE for stars satisfying this cut (black dashed in
the lower panel) is very similar to that of the total sample, with the
only noticeable difference being a suppression of high eccentricity
orbits in the first peak, not relevant for our analysis. Thus, in the
rest of the paper we use this restricted sample ($N=\num{22776}$
stars).

In order to check for the real significance of the peaks, as opposed
to random fluctuations, we perform the test proposed by
\cite{1981JRSSB..43...97S} -- see Appendix \ref{sec:silverman}. The
presence of (at least) three peaks is confirmed with high statistical
significance. This confirms the detection of three peaks by
\cite{2020MNRAS.492.3408P} with a restricted sample of 314 RR Lyrae
stars (under the assumption of Gaussian components, not assumed in our
analysis). This shows that this $\rper$ distribution is not associated
with any specific feature of their sample, but indicates the presence
of distinct Galactic components.

The red dashed curve in Fig.~\ref{fig:MW_e_vs_rper_all} (lower panel)
is the $\rper$ KDE obtained with the $h_\mathrm{crit}$ for three peaks
from Table \ref{tab:p_values}, i.e. the most detailed
KDE having no more than three peaks. The only, barely noticeable,
difference with respect to the black curve is the smoothing of the
little wiggles. The three peaks are located at
${\rper \approx 0.59\kpc}$, ${5.4\kpc}$ and ${7.14\kpc}$ (as indicated
by the vertical red dotted lines).

Next, we analyze the chemical and kinematic properties of stars in
these three populations and use the results to suggest their
association with known structures in the MW.

%%%%%%%%%%%%%%%%%%%%%%%%%%%%%%%%%%%%%%%%%%%%%%%%%%

\subsection{Halo, thick and thin discs}
\label{sec:halo_thin_thick}

\begin{figure}
  \includegraphics[width=\columnwidth]{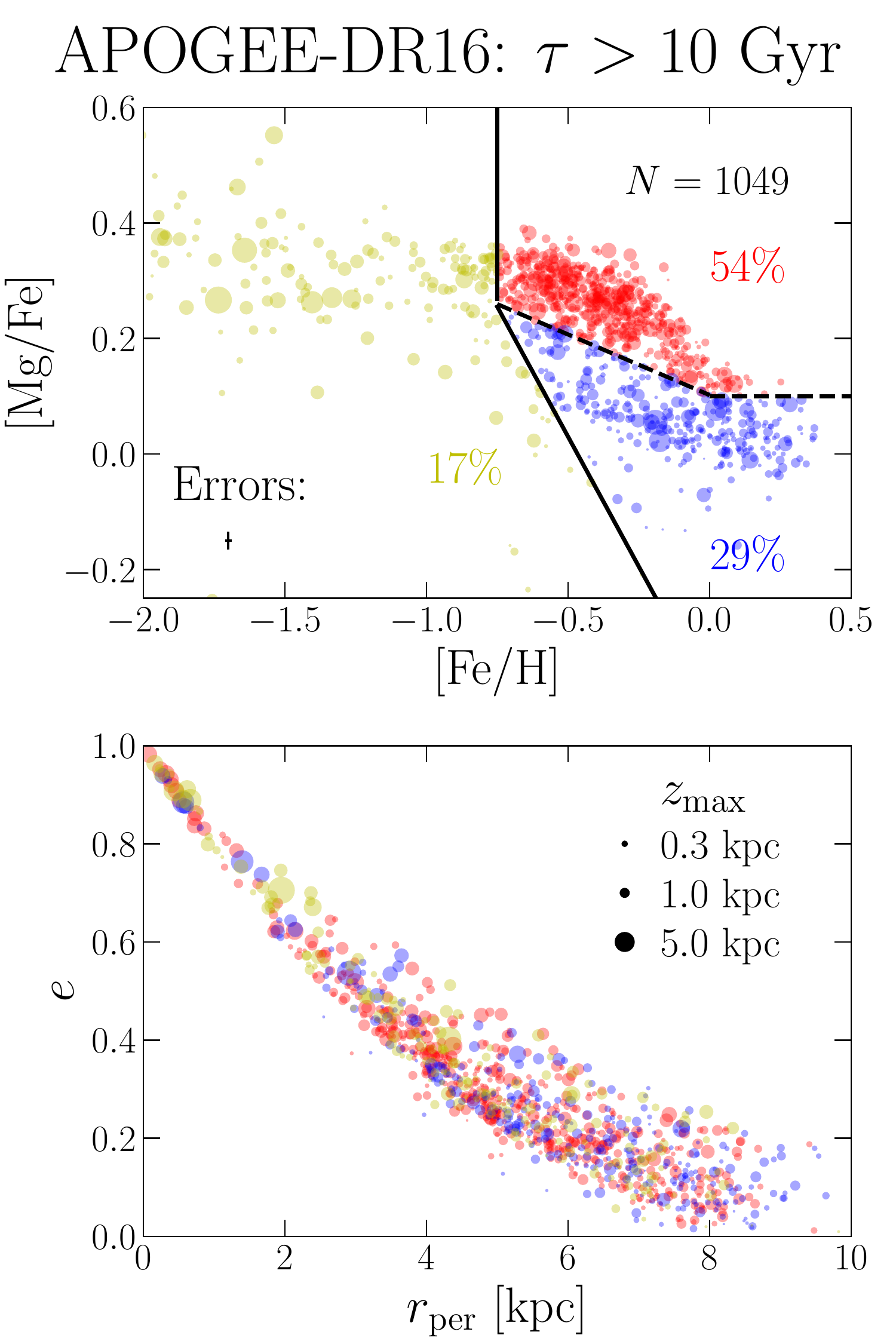}
  \caption{Chemical map (top) and eccentricity, $e$, vs pericenter
    radius, $\rper$, distribution (bottom), for the cross-match with
    APOGEE-DR16. Colours define halo (yellow), thick (red) and thin
    (blue) disc stars, split through the straight lines. The fractions
    of each component are shown in the top panel. Point sizes are
    proportional to $\zmax$. The black point in the top panel shows
    the median uncertainties ("errors") in [Fe/H] and
    [Mg/Fe]. Considering the peaks in Fig. \ref{fig:MW_e_vs_rper_all},
    the first (from left to right) is predominantly associated to the
    halo ($\alpha$-rich, very metal-poor and high $\zmax$), while the
    second and third have significant contributions from the thick
    disc ($\alpha$-rich, reasonably metal poor and intermediate
    $\zmax$) and the thin disc ($\alpha$-poor, metal-rich and small
    $\zmax$). The top panel suggests that the thin/thick discs were
    co-forming in the MW's very first Gyrs, with comparable star
    counts.}
\label{fig:apg_GMM_chem_kin_all}
\end{figure}

The catalogue of \cite{2018MNRAS.481.4093S} provides ages, 6-D phase
space coordinates and metallicities, but lacks information on the
$\alpha$-abundances. For this, we cross-match the sample with
APOGEE-DR16, which results in $1049$ (turnoff + giant) stars. The
chemical abundances ([Mg/Fe] and [Fe/H]) for these stars from
APOGEE-DR16 are shown in Fig. \ref{fig:apg_GMM_chem_kin_all} (upper
panel). Colours represent the different populations represented by the
separating straight lines \citep[defined by eye, similarly to the
definitions of e.g.][]{2011A&A...535L..11A, 2019MNRAS.482.3426M},
which define regions for the stellar halo (yellow), thick disc (red)
and thin disc (blue). The black point at the bottom left corner at the upper panel illustrates the median uncertainties in [Mg/Fe] and [Fe/H], as reported in APOGEE-DR16. The lower panel shows the distribution of these stars in the $e$ vs $\rper$ plot. In the two panels, point sizes are proportional to $\zmax$.

The most metal-poor and generally $\alpha$-rich stars (yellow points)
occupy the locus, in the chemical map, identified with the halo. These
stars are typically clustered around small $\rper$ and high $e$, and
have large vertical excursions, with $\zmax$ reaching $\gtrsim 10$
kpc, which is consistent with our assignment of these stars to the
stellar halo, with a probable accreted origin, as argued by
\cite{2018MNRAS.478..611B} and \cite{2018Natur.563...85H}. The red
points represent stars with typical thick disc chemical abundances,
i.e. they are $\alpha$-rich and reasonably metal-poor. These stars
spread across all values of $e$ and $\rper$ in the lower panel, but
seem to cluster around $\rper \approx 5\kpc$ and $e \approx
0.25$. Additionally, these stars have $\zmax$ as large as a few kpc,
which is also consistent with their classification as thick disc
stars. Finally, the $\alpha$-poor, metal-rich stars (blue points)
spread around $4\lesssim\rper/\kpc \lesssim 8$ and $e\approx 0.2$ and
typically have small vertical excursions $\zmax$, compatible with
their assignment to the thin disc.

At first sight the features just listed could suggest the
identification, from left to right in Fig. \ref{fig:MW_e_vs_rper_all},
of the first $\rper$ peak with halo stars, the second with the thick
disc and the third with the thin disc. However, a closer inspection
shows that both the thin and thick disc components contribute
distinctively to each of the second and third $\rper$ peaks, as we
show in Sec.~\ref{sec:dissec_thin_thick}. Whatever the case may be,
the identification of three peaks, seemingly associated to the halo,
thin and thick discs, agrees with the results from
\cite{2020MNRAS.492.3408P}, although the position of the second peak
differs from the value ${\rper\approx 3 \mathrm{kpc}}$ found in that
work. This reinforces their conclusion about the third peak detection
and its main relation to an old thin disc population, while showing
that it is not associated to any specific feature of their sample, but
it represents a significant Galactic component.

Our separation in the chemical space shown in
Fig.~\ref{fig:apg_GMM_chem_kin_all} results in $\sim 17\%$ of the
stars in the halo, $\sim 54\%$ in the thick disc and $\sim 29\%$ in
the thin disc. However, these fractions should not be considered as
precisely describing the whole Galaxy, since precise fractions would
require a detailed analysis of the APOGEE-DR16 selection function,
which is beyond the scope of this work. In Appendix \ref{sec:cuts} we
apply different simple geometric and age cuts and show that the
fractions of old thin and thick disc stars typically vary around
$\sim 20-30\%$ and $\sim 40-55\%$, respectively -- see
Table~\ref{tab:thin_thick_fractions}.

Most importantly, since [Fe/H] and [Mg/Fe] are conserved over a star's
life-time, Fig. \ref{fig:apg_GMM_chem_kin_all} shows that in the very first Gyrs of the MW's evolution, not only the thick disc but
{\it both} chemical thin and thick discs were forming, with comparable
star counts. The conclusion is similar, although with a poorer
statistics, if we select stars older than ${11\Gyr}$. In this case,
the cross-match with APOGEE-DR16 results in only 573 stars (with
median age uncertainty of ${\approx 0.84\Gyr}$). Of these, $21\%$
belong to the halo, $54\%$ to the thick disc and $25\%$ to the thin
disc -- see Table~\ref{tab:thin_thick_fractions}.  In
Sec.~\ref{sec:discuss}, we develop this conclusion and discuss the
implications for various thick disc formation scenarios.

We now focus on the thin and thick discs. When referring
to the $\rper$ peaks (Fig.~\ref{fig:MW_e_vs_rper_all}-lower panel), we neglect the left-hand peak
associated to the stellar halo, and call the remaining peaks, at
$\rper\approx 5.4\kpc$ and $\rper\approx 7.14\kpc$, the low and high
$\rper$ peaks, respectively.

%%%%%%%%%%%%%%%%%%%%%%%%%%%%%%%%%%%%%%%%%%%%%%%%%%

\subsection{Migrators and non-migrators in the thin and thick discs}
\label{sec:dissec_thin_thick}

\begin{figure*}
  \includegraphics[width=\textwidth]{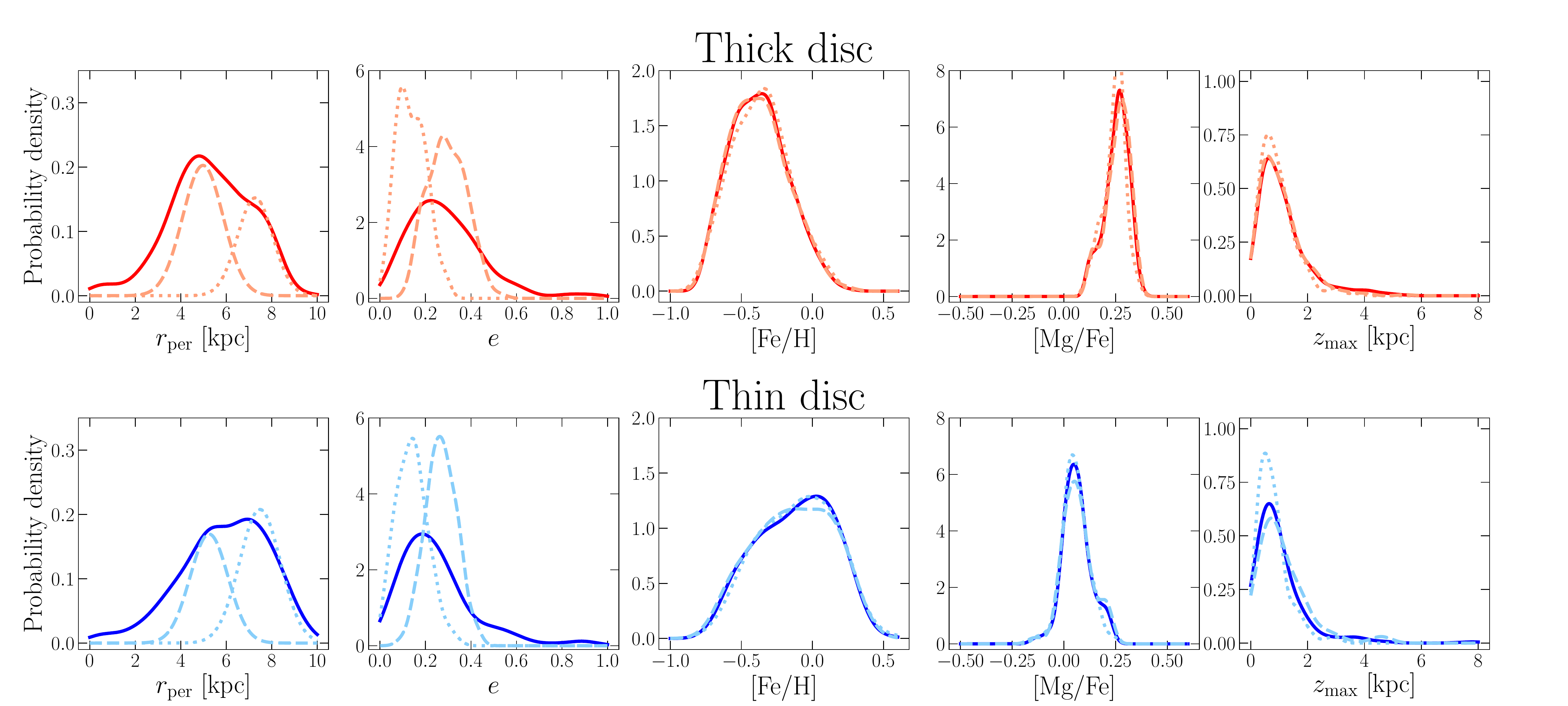}
  \caption{KDEs for kinematic quantities and chemical abundances for
    the thin disc (305 stars) and thick disc (570 stars) from
    APOGEE-DR16. The $\rper$ distribution has two peaks/bumps for both
    discs, which we model as two Gaussians, shown in the first column
    as dotted and dashed curves. In the other columns, dotted and
    dashed curves show the KDEs of the respective quantities for these
    two subsamples. The larger $\rper$ subsamples (dotted) are
    characterized by lower eccentricities while the lower $\rper$
    subsamples (dashed lines) by higher eccentricities. The chemical
    abundance distributions of these subsamples are very similar to
    that of the total sample inside each disc, suggesting that the
    $\rper$ peaks/bumps are not due to contamination from the
    thin/thick discs.}
  \label{fig:apg_chem_kin_kde}
\end{figure*}

The solid lines in Fig. \ref{fig:apg_chem_kin_kde} show the KDEs of
chemical and kinematic quantities for the chemically defined thick
disc (upper panel, 570 stars) and thin disc (lower panel, 305
stars). The thin disc $\rper$ distribution has two peaks, one at
$\rper\approx 5\kpc$ and another at ${\rper\approx 7\kpc}$, while the
thick disc peaks at ${\rper\approx 5\kpc}$ and has a bump at
${\rper\approx 7\kpc}$. This shows that the low and high $\rper$ peaks
(second and third, in Fig.~\ref{fig:MW_e_vs_rper_all}) have distinct
contributions from both the thin and thick discs, although the thick
disc contributes mostly to the low $\rper$ peak, while the thin disc
gives similar contributions to the low and high $\rper$ peaks. This
contrasts with the conclusions of \cite{2020MNRAS.492.3408P}, who
tentatively attribute the low and high $\rper$ peaks to the thick and
thin discs, respectively.

Using a Gaussian Mixture Model (GMM) \citep[][]{scikit-learn}, we
split the thin and thick discs $\rper$ distributions into three
Gaussians each, although only showing the two relevant components (the
third Gaussian, not shown, lies at small $\rper$ and large $e$ and is
probably due to contamination from the halo). These Gaussian
components are shown as dashed and dotted lines in the first column of
Fig. \ref{fig:apg_chem_kin_kde}. With these Gaussians, we define two
subsamples, according to the highest probability of belonging to one
or the other Gaussian component. In the other columns, the dashed and
dotted lines represent the KDEs of the corresponding quantities for
these fixed subsamples.

The second column shows that the thin disc eccentricity distribution
peaks at $e\approx 0.2$ and extends to $e\lesssim 0.7$, while the
thick disc peaks at $e\approx 0.25$ and is
slightly broader. For both discs, the high (low) $\rper$ subsample is
characterized by lower (higher) eccentricities. In principle, this
could be seen as a simple manifestation of the anti-correlation
between $\rper$ and eccentricity expressed by Eq.~(\ref{eq:1}) and
shown in Figs.~\ref{fig:MW_e_vs_rper_all} and
\ref{fig:apg_GMM_chem_kin_all}. However, the presence of the $\rper$
peaks and bumps (in both discs) indicates the existence of a distinct
population with $\rper\approx 7.1\kpc$, and these results show that
this population is characterized by nearly-circular orbits.

The KDEs for [Fe/H] and [Mg/Fe] (third and fourth columns) show that,
for both the thin and thick discs, the two Gaussian $\rper$ components
(dashed and dotted curves) have chemical distributions similar to
those of the total samples (solid lines), thus indicating that the
$\rper$ subsamples are typical chemical thin or thick disc samples,
and not due to cross contamination.

The simulation-based results of Sec.~\protect\ref{sec:time_evolution}
make clear that the high $\rper$ peak is produced by stars which
formed at inner radii and moved to the current location via radial
migration (churning), while the low $\rper$ peak is produced by
non-migrators. However, hints for this can be anticipated already
using simple arguments. For simplicity, we focus on the thin disc
(bottom row in Fig.~\ref{fig:apg_chem_kin_kde}): first, since the low
and high $\rper$ subsamples (dashed and dotted lines, respectively)
have similar ages ($\tau > 10\Gyr$), their similar chemical abundances
suggest that they must have been born at similar radii. Stars born at
the same radius but in orbits with different eccentricities would have
different $\rper$ but, instead of producing peaks or bumps (as
observed in Figs. \ref{fig:MW_e_vs_rper_all} and
\ref{fig:apg_chem_kin_kde}), a continuous (i.e. unimodal) distribution
would be expected in the $\rper$ distribution, if migration was not
taking place. Second, the metallicity distribution extends to (a
priori) high values for the outer radii, which was shown by
\cite{2016ApJ...818L...6L} to be produced by radial migration of stars
born at inner radii. Third, the high $\rper$ peak is associated with
stars in low-eccentricity orbits, for which radial migration is most
effective \citep[][]{2002MNRAS.336..785S,
  2012MNRAS.426.2089R}. Furthermore, radial migration does not heat
the orbits, i.e. it does not increase their eccentricities. This is
compatible with the idea that stars contributing to the high $\rper$
peak migrated to their current location while preserving low
eccentricities. Therefore, we anticipate the simulation-based results
of Sec.~\ref{sec:time_evolution} and call the stars composing the high
$\rper$ peak the {\it migrators}, and the ones composing the low
$\rper$ peak the {\it non-migrators}. In the GMM, $55\%$ of the thin
disc stars are classified as migrators, while this fraction decreases
to $38\%$ for thick disc stars -- see Fig.~\ref{fig:apg_chem_kin_kde}.

Finally, the right-hand column shows that the thin and thick discs
have similar $\zmax$ distributions, peaking at
${0.5\lesssim \zmax/\kpc\lesssim 1}$. Importantly, there is a large
fraction of chemical thick disc stars at small $\zmax$, i.e. the region associated
to the geometric thin disc. This large fraction is partially due to the adopted geometric cut (${d<2\kpc}$), which
naturally suppresses the high $\zmax$ portion of the thick
disc. However, in Appendix \ref{sec:cuts} we show that even selecting
stars with $|z|<6\kpc$, the fraction of chemical thick disc stars with
low $\zmax$ is still significant. Finally, it is interesting to note
that, for both the thin and thick discs, the migrator and non-migrator
subsamples have similar $\zmax$
distributions as the total samples.

%%%%%%%%%%%%%%%%%%%%%%%%%%%%%%%%%%%%%%%%%%%%%%%%%%

\section{$N$-body+SPH simulation of a clumpy galaxy}
\label{sec:sims}

We now explore an $N$-body+Smooth Particle Hydrodynamics (SPH)
simulation of the formation of an isolated galaxy which forms clumps
in its early evolution. In these clumps, the star formation rate
density is high and supernovae type II rapidly enrich the surrounding
medium with $\alpha$-elements. The clumps scatter $\alpha$-rich stars
to high $|z|$ ($\alpha$-poor stars are also scattered, but they are
born further the clumps and therefore are less likely to be strongly
scattered), giving rise to a geometric thick disc \citep[similar to
the mechanism proposed by][]{2009ApJ...707L...1B}. After
${\approx 4\,\mathrm{Gyr}}$, the clumps stop forming because of the
declining gas mass fraction and any remaining clumps are either
disrupted or sink to the centre; thus clumps only affect the early
evolution of the galaxy. This scenario was recently shown to
produce a chemical bimodality \citep[][]{2019MNRAS.484.3476C}, density
profiles \citep[][]{10.1093/mnras/staa065} and a Splash population
\citep[][]{2020ApJ...891L..30A} very similar to those observed in the
MW \citep[][]{2018AJ....156..125H, 2017MNRAS.471.3057M,
  2020MNRAS.494.3880B}. Finally, co-formation of the thin and thick
discs is a natural prediction of this scenario.

In this section, we first show that the clump scenario reproduces the
observational trends discussed above. Then, we analyze the
simulation's time evolution to unravel the secular evolution driving
the Milky Way to exhibit those trends.

%%%%%%%%%%%%%%%%%%%%%%%%%%%%%%%%%%%%%%%%%%%%%%%%%%

\subsection{Simulation details}

For the initial conditions, we use a $5\times$ higher mass resolution
version of that described in \cite{2019MNRAS.484.3476C}. A gas corona
with ${5 \times 10^6}$ particles in hydrostatic equilibrium is placed
within an NFW \citep{1997ApJ...490..493N} dark matter halo which
constitutes $90\%$ of the total mass. The virial radius and mass of
the dark matter halo are $r_{200} \simeq 200 \kpc$ and
$M_{200} = 10^{12} \Msun$, respectively, and it also has
$5 \times 10^6$ particles. While the gas corona has the same radial
density profile, it constitutes only $10\%$ of the mass. No additional
baryons or stars are present at $t=0$. The spin parameter of the gas
corona is set to $\lambda = 0.065$ \citep{2001ApJ...555..240B}. The
gas corona and the dark matter halo have softening lengths of
$\epsilon = 50 \pc$ and $\epsilon = 100 \pc$, respectively. Star
particles forming from the gas also have a softening of
$\epsilon = 50 \pc$.

These initial conditions are evolved for $13 \Gyr$ with {\sc gasoline}
\citep{gasoline, 2017MNRAS.471.2357W} with sub-grid models for the gas
cooling, star formation and feedback. Gas cooling includes the
metal-line cooling of \citet{2010MNRAS.407.1581S}. The gas cools and
settles into a disc. Once its density exceeds a certain threshold (set
to 0.1 cm$^{-3}$), star formation starts (with an efficiency set to
$5\%$) from gas particles with temperature below 15,000 K which are
part of a converging flow. We use the blast wave supernova feedback
\citep{2006MNRAS.373.1074S}, which couples, as thermal energy, $10\%$
of the $10^{51}$ erg per supernova to the interstellar medium. Gas
mixing uses turbulent diffusion as described by
\citet{2010MNRAS.407.1581S}. A base timestep of $\Delta t=5\Myr$ is
used, with timesteps refined such that
$\delta t = \Delta t/2^n < \eta\sqrt{\epsilon/a_g}$, where we set the
refinement parameter $\eta = 0.175$. We set the opening angle (for the
tree-code gravity calculation) to $\theta = 0.7$. The timesteps of gas
particles additionally satisfy the condition
$\delta t_{gas} = \eta_{courant}h/[(1 + \alpha)c + \beta \mu_{max}]$,
where $\eta_{courant} = 0.4$, $h$ is the SPH smoothing length set over
the 32 nearest neighbour particles, $\alpha$ and $\beta$ are the
linear and quadratic viscosity coefficients and $\mu_{max}$ is
described by \citet{gasoline}.

Oxygen and iron yields from SNII and SNIa are taken from
\cite{1996A&A...315..105R}. As in \cite{1996A&A...315..105R}, we use
Padova stellar lifetimes to determine SNII rates, while SNIa rates are
determined from those same lifetimes in a binary evolution model.

We use data from the last snapshot, at $13\Gyr$, and select star
particles older than $10 \Gyr$ at the Solar torus, i.e. within a torus
defined by circles at ${R = 8\pm 2\kpc}$, for an approximate match
with the selection criteria used for the MW. For brevity, we will
refer to this region as the Solar Neighbourhood in the simulation.  We
compute the total gravitational potential at this final snapshot and
integrate each orbit for 10 dynamical times using {\sc agama}
\citep[][]{2019MNRAS.482.1525V}. Finally, we calculate pericenter and
apocenter radii, as well as eccentricities, using Eq.~(\ref{eq:1}),
and $\zmax$.

%%%%%%%%%%%%%%%%%%%%%%%%%%%%%%%%%%%%%%%%%%%%%%%%%%

\subsection{Simulation results and comparison to the MW}

\begin{figure}
  \includegraphics[width=\columnwidth]{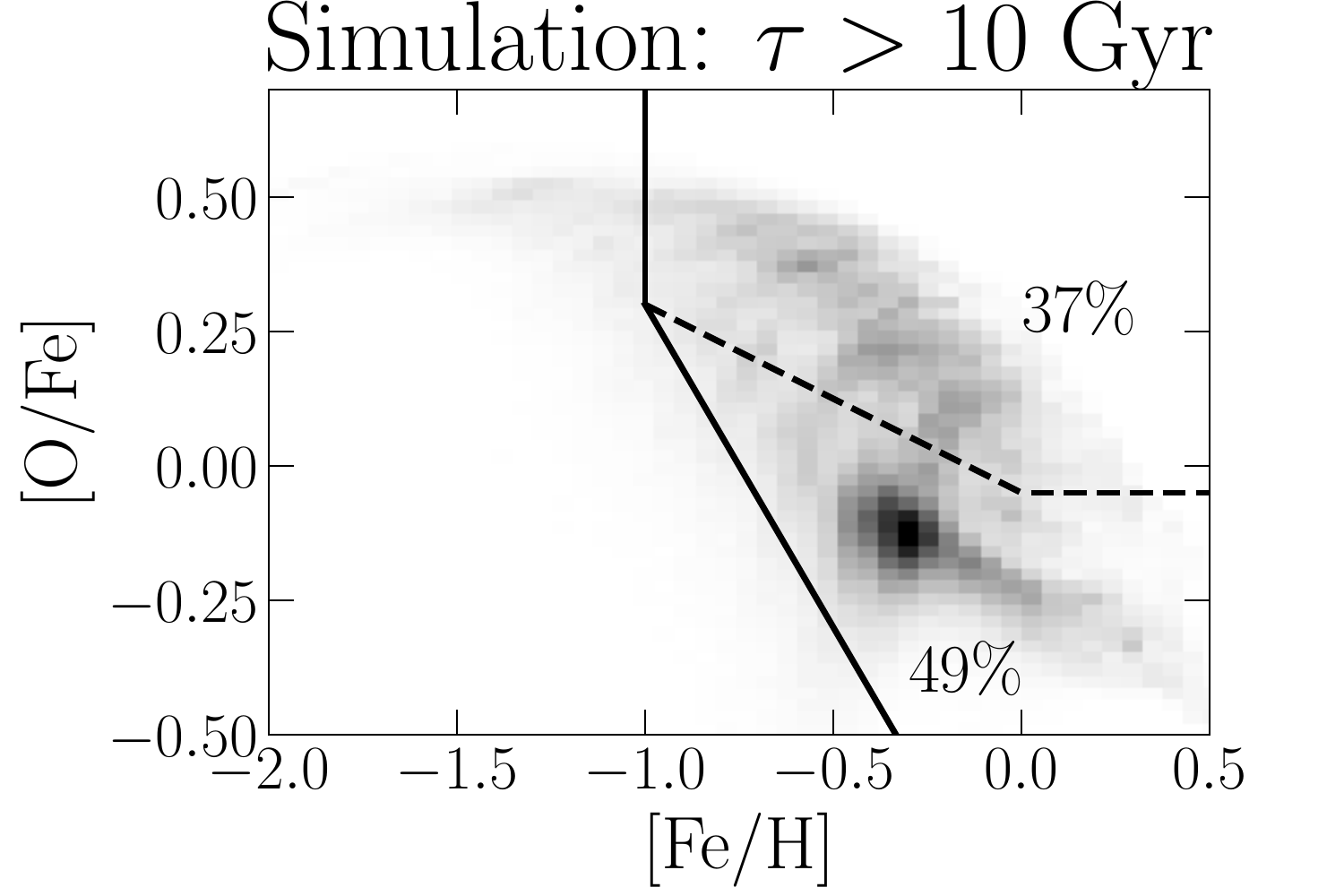}
  \caption{2D histogram in chemical space for old stars in the
    simulation. Like in the MW (Fig.~\ref{fig:apg_GMM_chem_kin_all}),
    a chemical bi-modality is already present in the very first Gyrs
    of evolution. Dashed lines define the chemical thick and thin
    discs.}
  \label{fig:sim_old_chem_map}
\end{figure}

Fig.~\ref{fig:sim_old_chem_map} shows a 2D histogram in the abundance
map ([O/Fe] vs [Fe/H]) for old star particles ($\tau > 10$ Gyr)
produced in the simulation. The immediate conclusion is that, as in
the MW, the simulated galaxy develops the chemical bi-modality
associated to the thick/thin discs in the very first Gyrs of its
evolution, indicating a rapid formation of the thick and thin
discs. The straight lines in this plot define chemical thick and thin
discs (above and below the dashed lines, respectively), similar to
those defined in the MW (Fig.~\ref{fig:apg_GMM_chem_kin_all}-upper
panel). The thick disc comprises $37\%$ and the thin disc, $49\%$ of
the total. If we instead select star particles older than $11$ Gyr, we
have $49\%$ for thick disc and $34\%$ for thin disc. These numbers do
not match in detail the fractions in the cross-match with APOGEE-DR16
(Fig.~\ref{fig:apg_GMM_chem_kin_all} and Table
~\ref{tab:thin_thick_fractions}). However, precise comparisons would
require detailed modelling of the APOGEE-DR16 selection function,
which is beyond the scope of this work.

\begin{figure}
  \center
  \includegraphics[width=\columnwidth]{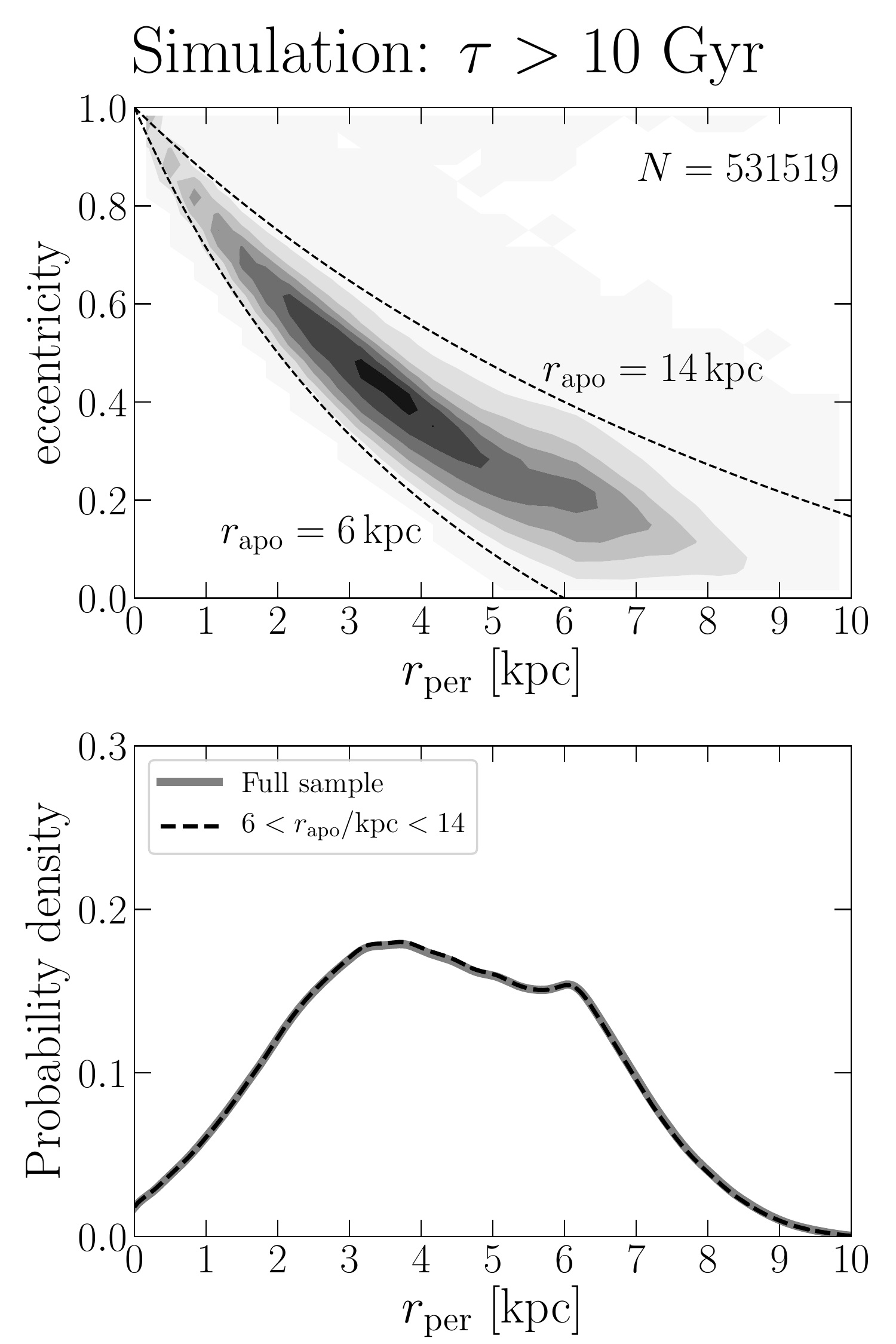}
  \caption{Top: distribution of old stars in eccentricity vs
    pericentric radius for the simulation Solar Neighbourhood. Bottom:
    KDEs of $\rper$: for the full sample (grey) and for
    $6 < \rper/\mathrm{kpc} < 14$ (black dashed). As in the MW
    (Fig.~\ref{fig:MW_e_vs_rper_all}), we identify two peaks, mainly
    (but not only) associated with the thick and thin disc, from left
    to right. The peak associated to the accreted halo in the MW is
    not present, since the simulated galaxy evolves in isolation.}
  \label{fig:sim_e_vs_rper}
\end{figure}

Fig.~\ref{fig:sim_e_vs_rper} (upper panel) shows the distribution of
stars in the $\rper$ vs $e$ plane of the simulation ($N=\num{531519}$
star particles). As in the MW (Fig. \ref{fig:MW_e_vs_rper_all}), the
data peak along a line of approximately constant $\rapo$, at
$8 \lesssim \rapo/\mathrm{kpc}\lesssim 9$, and is mostly concentrated
in the region $6 \lesssim \rapo/\mathrm{kpc}\lesssim 14$ (dashed
lines), reflecting the same selection effects present in our MW
sample, as described in Sec.~\ref{sec:MW_peaks}. The bottom panel
shows the $\rper$ KDE for this sample (full grey) and after the
restriction to the region $6 \lesssim \rapo/\mathrm{kpc}\lesssim 14$
(dashed black), which reduces the sample to $N=\num{513907}$ star
particles. This restriction leaves the KDE essentialy
identical to the original one. We use this restriction in the rest of
the analysis for the same reason as we did with the MW data, i.e. to
avoid orbits with extreme or unrealistic kinematic properties, which
can be introduced by the assumed potential.

The simulated galaxy lacks the first peak at ${\rper\approx 0.5\kpc}$,
which in the MW is associated with the accreted halo; this
identification is strengthened by the fact that the simulated galaxy
evolves in isolation, with no accretion events.  The two other MW
peaks are present in the simulation, at $\rper \approx 3.5\kpc$ and
$\rper \approx 6\kpc$. The positions of these peaks differ from those
in the MW (Fig.~\ref{fig:MW_e_vs_rper_all}), as well as the
eccentricity values at the peaks. However, these differences are not
fundamentally important for our conclusions, since we are interested
in comparing the trends, while the simulated galaxy is not intended to
be an exact replica of the MW. Next, we show that indeed these two
peaks are mostly, but not only, composed of thick and thin disc stars,
respectively. For a better comparison with the MW, in what follows we
exclude the most metal-poor star-particles (those to the left of the
solid line in Fig.~\ref{fig:sim_old_chem_map}).

\begin{figure*}
  \includegraphics[scale=0.35]{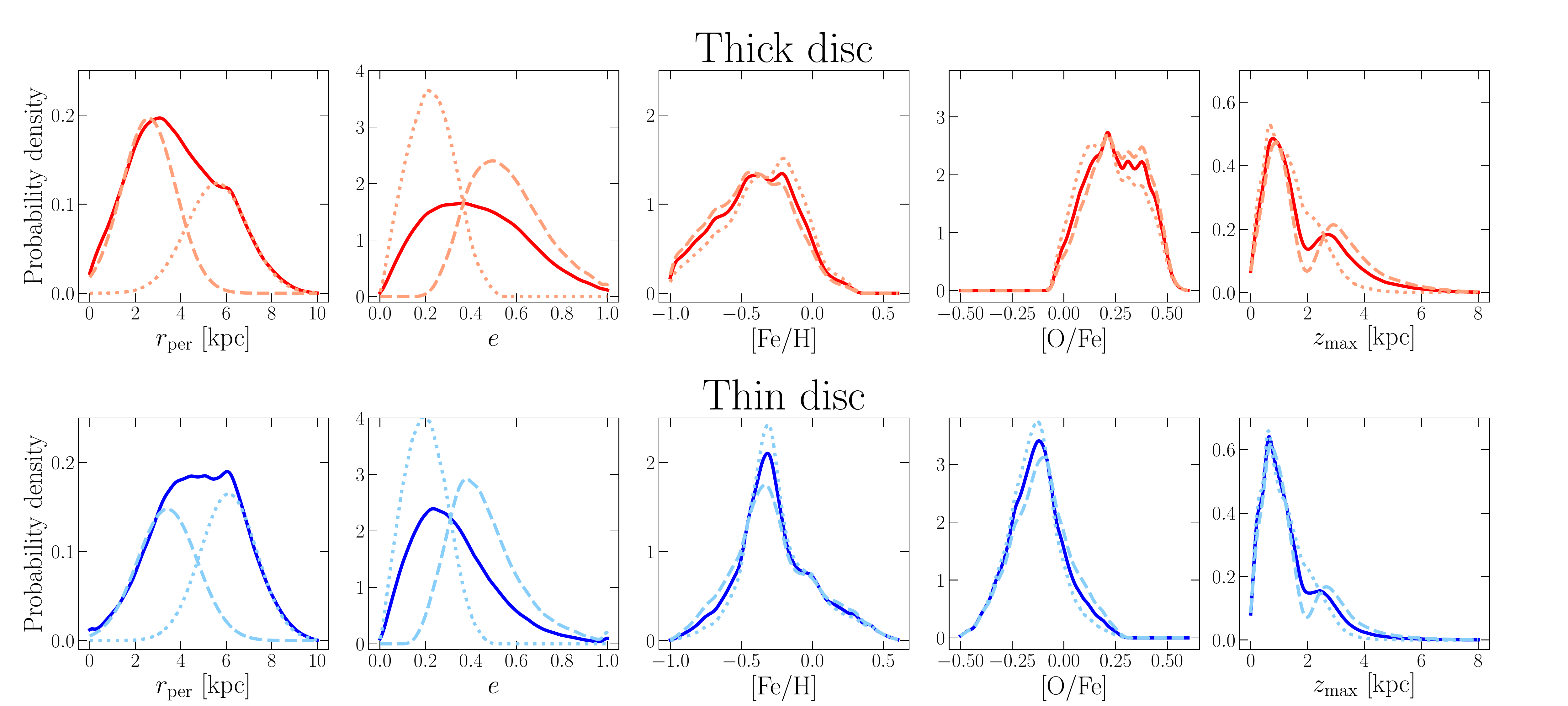}
  \caption{KDEs for kinematic quantities and chemical abundances for
    the chemical thin and thick discs in the simulation. As in the MW
    (Fig.~\ref{fig:apg_chem_kin_kde}), both discs have two peaks/bumps
    in the $\rper$ distribution, split into the two Gaussians shown as
    dotted and dashed curves in the first column. In other columns,
    dotted and dashed curves show the KDEs of the respective
    quantities for these two subsamples. The larger $\rper$ subsample
    (dotted) is characterized by lower eccentricities while the lower
    $\rper$ subsample (dashed) is characterized by higher
    eccentricities. As in the MW, the chemical abundance distributions
    of these subsamples are very similar to that of the total sample
    for each disc, suggesting that the $\rper$ peaks/bumps are not due
    to contamination from the thin/thick discs. The relative
    amplitudes of the two $\rper$ peaks/bumps in the thin and thick
    discs are similar those observed in the MW -- see
    Fig.~\ref{fig:apg_chem_kin_kde}.}
  \label{fig:sim_old_GMM_chem_kin_kde}
\end{figure*}

Fig.~\ref{fig:sim_old_GMM_chem_kin_kde} shows the KDE of the kinematic
quantities and chemical abundances obtained in the simulation, for the
thick disc (upper panels) and thin disc (bottom panels), to be
compared to Fig.~\ref{fig:apg_chem_kin_kde}. The $\rper$ distributions
of both thin and thick discs are bi-modal: the thick disc peaks at
$\rper\approx 3\kpc$ and has a bump at $\rper\approx 6\kpc$, while the
thin disc has a peak at $\rper\approx 4\kpc$ and another, of slightly
larger amplitude, at $\rper\approx 6\kpc$. The relative amplitudes of
these peaks and bumps agree well with those seen in the MW
(Fig.~\ref{fig:apg_chem_kin_kde}).

We repeat the procedure used for the MW and use a GMM to split the
$\rper$ distributions of the thin and thick discs into two Gaussians
each (dashed and dotted curves in the left panels). Then, for
subsamples defined according to the highest probability of belonging
to one or the other Gaussian component, we calculate the KDEs of other
quantities, shown in the respective panels of
Fig.~\ref{fig:sim_old_GMM_chem_kin_kde}. The eccentricities in the
thin disc peak at $e\approx 0.2$ and have a narrower distribution than
the thick disc, which peaks at $e\approx 0.35$. However, the thin disc
has a tail for large eccentricities and both distributions are broader
than those in the MW (Fig. \ref{fig:apg_chem_kin_kde}-second column).

The [Fe/H] and [O/Fe] KDEs are shown in the third and fourth columns.
The subsamples (dotted and dashed lines) have KDEs very similar to the
total samples, indicating that, as in the MW, the peaks and bumps in
the $\rper$ distribution are not due to contamination between the thin
and thick discs. Finally, the right-hand column shows the $\zmax$
distribution. The thick disc distribution is broader than that of the
thin disc. It is interesting to note that the simulation $\zmax$ KDE
is bi-modal, while this bi-modality is apparently not present in
Fig.~\ref{fig:apg_chem_kin_kde}. However, this bi-modality, associated
to discrete tracks in a plot of $\zmax$ vs $\rapo$, has been observed
in larger samples in the MW \citep[][]{2018ApJ...863..113H}, and shown
to result from different orbital families in the adopted axisymmetric
potential model -- see \protect\cite{2020MNRAS.492.3816A,
  2020arXiv200904849K}.

Thus, we conclude that the simulation reproduces the
features observed for the old stars in the MW, mainly:
\begin{itemize}
\item a chemical bi-modality comprising the thin and
  thick discs is already present in the very first Gyrs of evolution, with relative contributions comparable to those in the MW;
\item the $\rper$ distribution has two peaks that can be mainly
  attributed to the thin and thick discs, although both peaks have
  contributions from both discs and this is not due to contamination;
\item $\sim 1/2$ of the thin disc stars are associated to the high
  $\rper$ peak (at $\approx7$ kpc in the MW and at $\approx 6$ kpc in
  the simulation), while that fraction is $\sim 1/3$ for the thick
  disc;
  
\item stars that mostly contribute to the high $\rper$ peak are those
  with smaller eccentricities; and finally,
\item a significant fraction of chemical thick disc stars have low
  $\zmax$, i.e. overlap with the geometric thin disc.
\end{itemize}

We now analyze the simulation evolution to show that stars at the high
$\rper$ peak formed at smaller radii and moved outwards due to radial
migration, suggesting that the same happened in the MW.

%%%%%%%%%%%%%%%%%%%%%%%%%%%%%%%%%%%%%%%%%%%%%%%%%%

\subsection{Time evolution and radial migration}
\label{sec:time_evolution}

\subsubsection{Migrators all over the disc}
\label{sec:migration_all_disc}

We start this section by comparing the current position (at
${t=13 \Gyr}$) of old star particles on nearly-circular orbits with
the position they had at ${t=5 \Gyr}$. This earliest snapshot is
chosen right after the early turbulent period when the clumps are
still present, since we are now interested in the subsequent secular
evolution. We select star particles older than $10\Gyr$, but without
any geometric cut. For each particle we calculate the circularity
$\lambda_c = L_z/L_c(E)$, where $L_z$ is the angular momentum
$z$-component and $L_c(E)$ is the angular momentum of a circular orbit
at the same energy $E$. Then, we select particles on nearly-circular
orbits ($\lambda_c>0.9$) and split this sample into chemical thin and
thick discs (as defined in Fig.~\ref{fig:sim_old_chem_map}).

\begin{figure*}
  \center
  \includegraphics[scale=0.35]{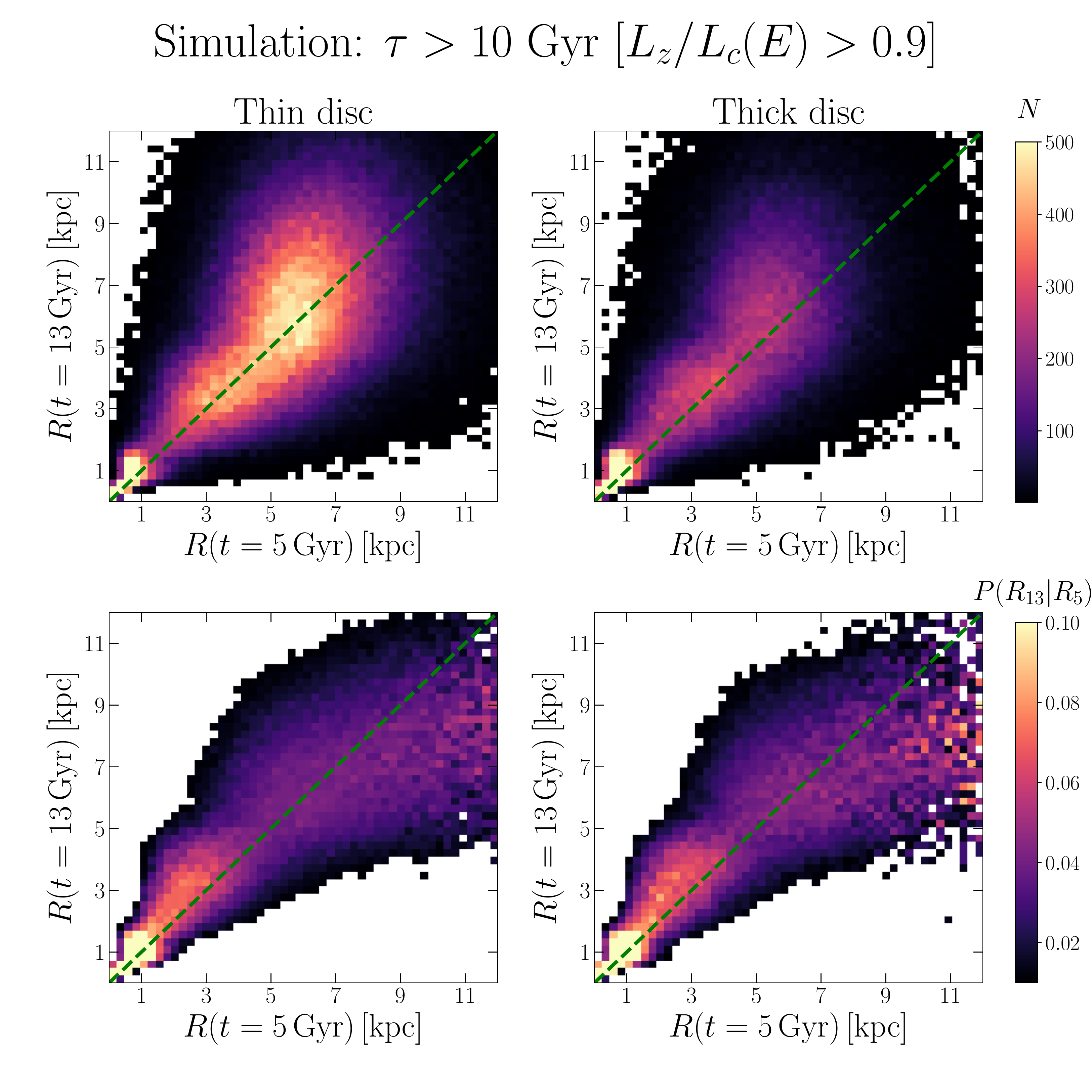}
  \caption{2D histograms of (cylindrical) radii $R(t=13\Gyr)$ vs
    ${R(t=5\Gyr)}$ for old star particles on nearly-circular orbits,
    split into thin/thick discs (left/right). Top panels are
    colour-coded by total numbers, while bottom panels are
    column-normalized. The large spread around the diagonal shows the
    effect of radial migration (churning) produced by spiral
    arms. This affects the thin and thick discs similarly, although
    particles on nearly-circular orbits (those most affected) are less
    numerous in the thick disc.}
  \label{fig:sim_old_R_5_13}
\end{figure*}

Fig.~\ref{fig:sim_old_R_5_13} shows 2D histograms of the cylindrical
radius at different times, ${R(t=13\Gyr)}$ vs ${R(t=5\Gyr)}$ for this
sample, split into thin (left) and thick (right) discs. Colours in the
top row represent total numbers, while the bottom row histograms are
column-normalized. If particles kept on orbits at the same early radii
they had at early times, they would concentrate along the diagonal
dashed line. However, a large spread is observed, with many particles
changing radii by several \kpc. Since we select particles on
nearly-circular orbits, we suppress the possibility of migration by
blurring, i.e. by oscillations around the guiding radii. Thus,
Fig.~\ref{fig:sim_old_R_5_13} shows the effect of radial migration
(churning) all over the disc, i.e. particles moving from
nearly-circular orbits to other nearly-circular orbits, as proposed by
\cite{2002MNRAS.336..785S}. Importantly, although the number of
nearly-circular orbits in the thick disc is smaller than in the thin
disc, as shown in the top panels, those orbits are similarly affected
by radial migration in the two discs, as shown in the bottom panels.

\begin{figure}
\begin{center}
  \includegraphics[scale=0.4]{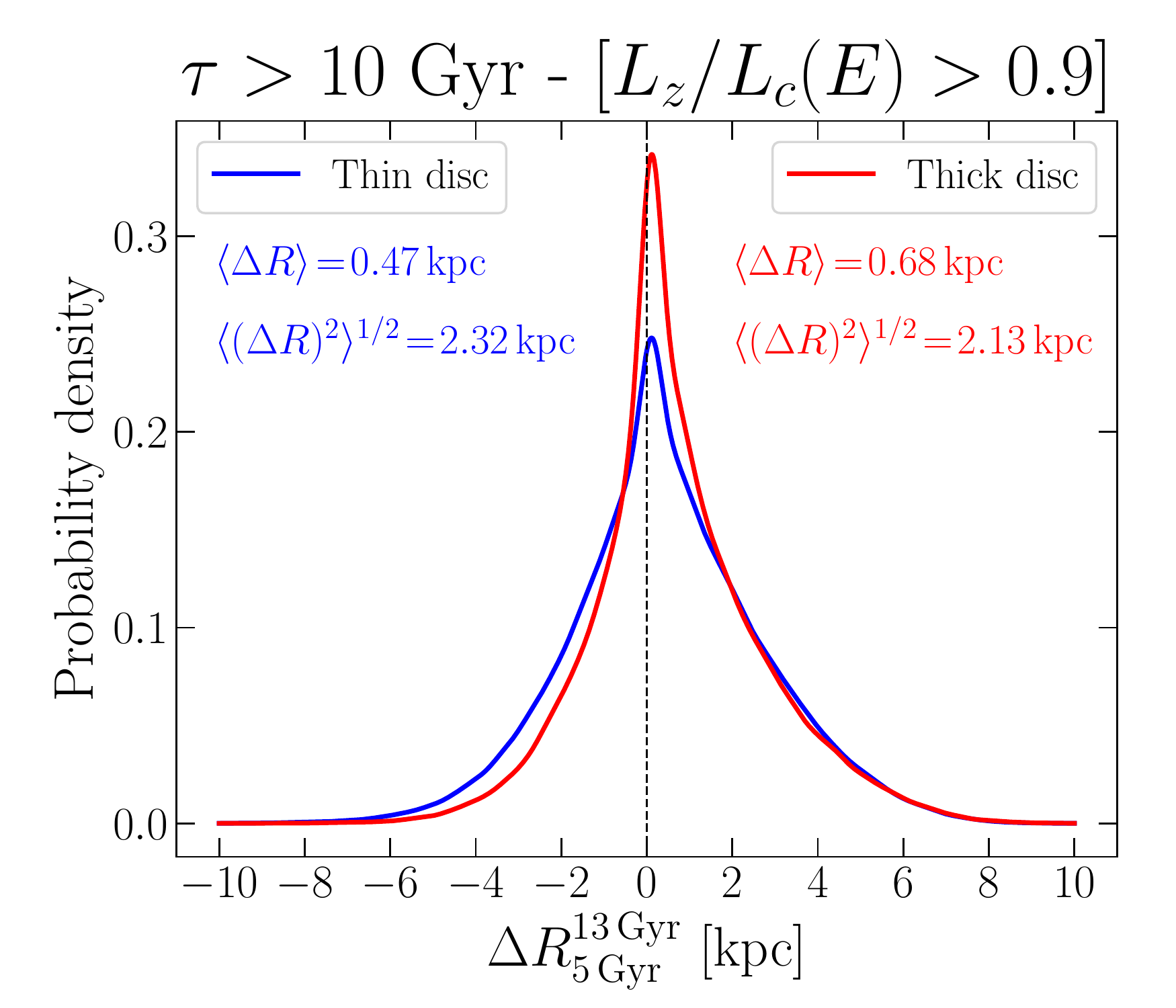}
  \caption{KDEs of $\Delta R \equiv R(t=13\Gyr) - R(t=5\Gyr)$ for the
    thin (blue) and thick (red) discs. Both KDEs peak near
    $\Delta R=0$, but have long tails, with star particles migrating
    by several \kpc. The positive skewness indicate a net outward
    movement.}
  \label{fig:sim_old_kde_delta_R}
 \end{center}
\end{figure}

For a compact representation of the effect of radial migration, we
calculate ${\Delta R \equiv R(13\Gyr) - R(5\Gyr)}$ for star particles
on nearly-circular orbits ($\lambda_c>0.9$) and estimate the KDE of
$\Delta R$. This is shown in Fig.~\ref{fig:sim_old_kde_delta_R}, with
the blue (red) curve representing the thin (thick) disc. The two discs
are similarly affected by radial migration: both KDEs have a narrow
peak near $\Delta R = 0$, and are positively skewed, i.e. have a
longer tail to the right, signing a net movement outwards, a
consequence of the density decrease outwards. The radius change
root-mean-square (rms) is $\sim 2.3\kpc$ for the thin disc
\citep[similar to that found by][]{2008ApJ...675L..65R} and
$\sim 2.1\kpc$ for the thick disc. This can be contrasted with the
results of \citet{2020ApJ...896...15F}, who found
${\langle (\Delta R)^2\rangle^{1/2} \sim 2.6\kpc\sqrt{\tau/6\Gyr}}$
for MW thin disc stars spanning ${5 \lesssim R/\kpc \lesssim 13}$.
For ${\tau = 8\Gyr}$ (the time between the two snapshots we are
considering), this results in
${\langle (\Delta R)^2\rangle^{1/2} \sim 3\kpc}$, while applying the
same geometric cut we find
${\langle (\Delta R)^2\rangle^{1/2} \sim 2.6\kpc}$ for the thin
disc. Therefore, the extent of migration in the simulation is
comparable to that in the MW.

\subsubsection{Migrators in the Solar Neighbourhood}
\label{sec:migration_solar_neig}

Having shown the effect of radial migration all over the thin and
thick discs, we now investigate its role in producing the high $\rper$
peak shown in Figs.~\ref{fig:MW_e_vs_rper_all},
\ref{fig:apg_chem_kin_kde}, \ref{fig:sim_e_vs_rper} and
\ref{fig:sim_old_GMM_chem_kin_kde}. We go back to the original
geometric cut and select, in the final snapshot ($t=13\Gyr$), star
particles older than $10\Gyr$ in the Solar Neighbourhood (a torus
defined by circles at ${R = 8\pm 2\kpc}$). Then, we track these same
particles at different snapshots from $5\Gyr$ to $13\Gyr$. At each
snapshot, we integrate the particles orbits in the respective
potentials and determine the peri and apo-center radii and
eccentricity for each particle.

\begin{figure*}
  \includegraphics[scale=0.35]{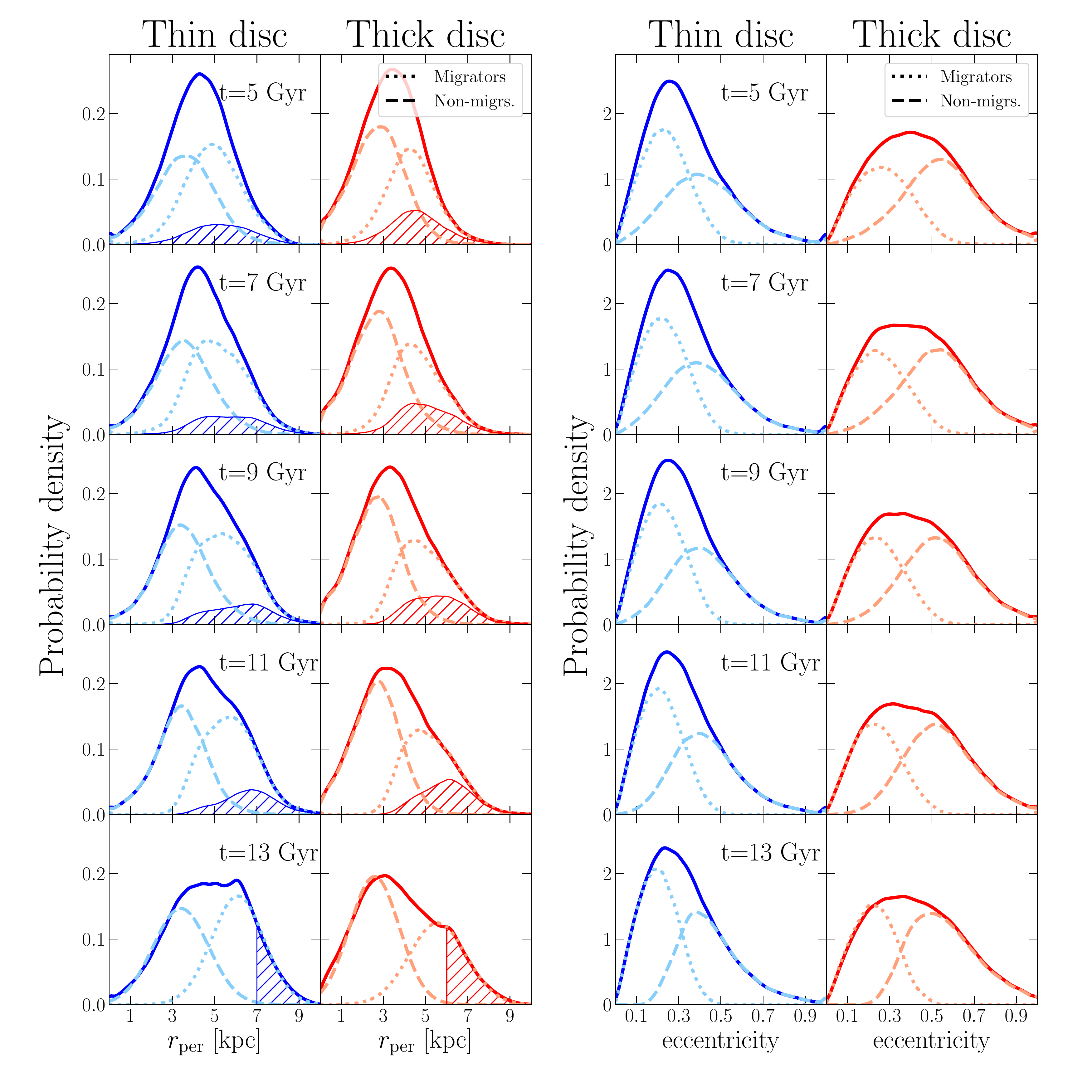}
  \caption{Left: time evolution of the $\rper$ KDE in the thin (blue)
    and thick (red) discs of the simulation. Old star-particles
    ($\tau > 10\Gyr$) are selected and split into two Gaussians at
    $t=13 \Gyr$. These fixed samples and subsamples are analyzed at
    earlier times (same particles considered in all snapshots). Right:
    time evolution of the eccentricities KDE in the thin and thick
    discs for the fixed samples defined in the left panels.  In both
    thin and thick discs, the low $\rper$ subsample (dashed) has
    higher eccentricities, and thus is less affected by radial
    migration, staying nearly fixed in place (the non-migrators). The
    high $\rper$ subsample (dotted) is that on nearly-circular orbits,
    and thus is significantly affected by radial migration and tends
    to move outwards (the migrators), developing a positive
    skewness. At $t=13\Gyr$, the migrators produce the high $\rper$
    peaks. Hatched areas show samples selected at $t=13 \Gyr$ as those
    with ${\rper > 7\kpc}$ (${\rper > 6\kpc}$) for the thin (thick)
    disc, showing the effect in a sample less contaminated by
    non-migrators. The migrators' eccentricity distributions are
    conserved, agreeing with expectations for radial
    migration. Non-migrators exhibit some heating, with
    nearly-circular orbits at $t=5 \Gyr$ getting more eccentric with
    time.}
  \label{fig:sim_old_rper_ecc_kde_tsteps}
\end{figure*}

Fig.~\ref{fig:sim_old_rper_ecc_kde_tsteps} (two left columns) shows
the $\rper$ KDEs (solid lines) for the thin (blue) and thick disc
(red) at different snapshots (rows). Similarly to
Figs.~\ref{fig:apg_chem_kin_kde} and
\ref{fig:sim_old_GMM_chem_kin_kde}, at the final snapshot (bottom
panels), we split our $\rper$ sample into two Gaussian components
using a GMM (dashed and dotted curves in the bottom panels) and use
these components to define two subsamples. Then, for these fixed
subsamples we calculate, at each snapshot, the $\rper$ KDEs, shown as
dotted and dashed curves in the remaining plots (weighted according to
the Gaussian weights obtained at the final snapshot).

At $5 \Gyr$, the full sample
distributions (thick solid lines) are unimodal for both thin and thick discs, peaking at
${\rper\approx 4\kpc}$ for the thin disc and at ${\rper\approx 3\kpc}$
for the thick disc, which must reflect differences in average
formation radii for these two components. From top to bottom panels,
we see that while the positions of these peaks are approximately
conserved, the distributions get progressively positively skewed,
until another peak appears at ${\rper\approx 6 \kpc}$ for the thin
disc and a bump at this same location for the thick disc.

For both thin and thick discs, while the low $\rper$ Gaussian
component (dashed) stays approximately fixed at the same position for
different snapshots, the high $\rper$ component (dotted) moves to the
right and becomes skewed to large values, producing the high $\rper$
peak at the final snapshot. This shows that the high $\rper$ peak is
characterized by stars which, on average, formed in inner radii and
move outwards. The hashed regions in
Fig.~\ref{fig:sim_old_rper_ecc_kde_tsteps} are defined as those
star-particles which, at the final time, have ${\rper > 7\kpc}$
(${\rper > 6\kpc}$) for the thin (thick) disc. Beyond these values,
the contribution from the low $\rper$ peak Gaussian component is
negligible, allowing to show the effect of migration in a cleaner
way. Some star-particles move at least $4 \kpc$ outwards in 2 Gyrs
(from 11 Gyr to 13 Gyr).

Besides the general tendency of a net outward movement observed in
Sec.~\ref{sec:migration_all_disc}, this is a consequence of the
inside-out growth of the galactic disc and the selection of old stars
in the Solar Neighbourhood, which was poorly populated at early
stages. Stars can visit the Solar Neighbourhood by oscillating around
their guiding radius (blurring, most relevant for eccentric orbits) or
via radial migration (churning). As mentioned in
Sec.~\ref{sec:dissec_thin_thick}, radial migration is more effective
for low-eccentricity orbits and it does not significantly change these
eccentricities. In other words, radial migration (churning) typically
moves stars from circular orbits to other circular orbits
\citep[][]{2002MNRAS.336..785S, 2012MNRAS.426.2089R}. In
Fig.~\ref{fig:sim_old_GMM_chem_kin_kde}, we showed that the high
$\rper$ component is the one with the lower eccentricities, in
agreement with the MW (Fig.~\ref{fig:apg_chem_kin_kde}). Moreover,
Fig. \ref{fig:sim_old_rper_ecc_kde_tsteps} (two right columns) shows
the KDEs of the eccentricity at the same times and for the same
samples as in the left panels. The eccentricities for those stars
which will produce the high $\rper$ peak in the left panels (dotted
lines) do not show a significant evolution. On the other hand, the
other subsample, which will produce the low $\rper$ peak in the two
left columns (dashed lines), shows some evolution, moving the
left-hand tail to the right, i.e. with any low-eccentricity orbits
gradually becoming more eccentric (heating).

Therefore, these results strongly suggest that the process responsible
for moving a population outwards and producing the high $\rper$ peak
(in both the simulation and in the MW) is radial migration (churning),
which justifies calling this population `the migrators' (and calling
the low $\rper$ peak `the non-migrators').

%%%%%%%%%%%%%%%%%%%%%%%%%%%%%%%%%%%%%%%%%%%%%%%%%%
%%%%%%%%%%%%%%%%%%%%%%%%%%%%%%%%%%%%%%%%%%%%%%%%%%

\section{Discussion}
\label{sec:discuss}
\subsection{Implications for the disc formation}
\label{sec:implications_disc_formation}

As mentioned in the Introduction, some theoretical models predict a
sequential formation of thin and thick discs, with the thin disc
becoming a significant component only after the thick disc has
formed. In such scenarios, one should not expect to find a significant
population of old thin disc stars. In Sec.~\ref{sec:results} we showed
several lines of evidence for such a population in the MW (Figs.~\ref{fig:apg_GMM_chem_kin_all} and \ref{fig:apg_chem_kin_kde}),
with thin (thick) disc stars comprising $29\%$ ($54\%$) of the
sample.

The existence of a significant old thin disc population suggests that
the thin disc starts forming earlier than generally assumed, having a
large formation time overlap with the thick disc, and this is the main
result of this work. Therefore, models which predict a sequential
formation of the thin/thick discs, such as the upside-down scenario,
are disfavored by our results. Also disfavoring this scenario is our
finding of a large fraction of chemical thick disc stars with low
$\zmax$ (peaking at $\zmax\approx 1\kpc$ -- see
Fig.~\ref{fig:apg_old_zmax_kde}), which suggests two possible
explanations: either, in line with the upside-down scenario, these low
$\zmax$ stars were born at high $|z|$ and settled to the geometric
thin disc, or the chemical thick disc stars were born at low $|z|$ and
a fraction of these stars were scattered to high $|z|$, leaving a
significant fraction at low $\zmax$. Since stars constitute a
collisionless component, one expects them to get kinematically hotter
under secular evolution (unless some fine-tuned mechanism is advocated
to kinematically cool these stars), thus making the second possibility
more plausible. Recently, \cite{2020arXiv200303316C} estimated ages
from APOGEE-DR14 survey data using a Bayesian Machine Learning
framework. Their results indicate that, while the inner and outer
discs seem to have different chemical evolutionary pathways, there is
a smooth transition between the formation of the thick disc and the
inner thin disc \citep[see also][]{2019MNRAS.489.1742F}. A significant
time-overlap in the formation of these components is also apparent in
their results (although these authors did not explore this topic
further), which are thus complementary to the results found by us.

On the other hand, some models do predict an early co-formation of
thin and thick discs and thus the existence of a conspicuous old thin
disc population. In the accretion model, \cite{2003ApJ...597...21A}
predict that $15\%$ of the current thin disc stars should be older
than ${10\Gyr}$. In this picture an old thin disc is built up from
several disrupted satellites, each with unique chemical signatures,
geometry and kinematics. Thus, \cite{2003ApJ...597...21A} predicted
that this old thin disc should show distinct chemo-kinematic
correlations. In our sample, the subsamples of stars classified as
migrators and non-migrators (based on $\rper$, which is similar to
splitting by eccentricity) have chemical abundance distributions very
similar to the total samples --
(Fig.~\ref{fig:apg_chem_kin_kde}). This suggests that the correlations
predicted in the accretion scenario are not observed in the MW.

In the major merger scenario \citep{1993ApJ...403...74Q,
  2008MNRAS.391.1806V, 2018Natur.563...85H}, the thick disc is
produced by a merger whose progenitor is massive enough to vertically
heat a proto-disc, and an old thin disc remnant is
expected. \cite{2008MNRAS.391.1806V} predict that the mass fraction of
this remnant thin disc is $15\%-25\%$ of the total stellar mass at the
end of the merger. Note that this refers to a geometric definition,
while the chemical thin disc is generally seen as forming only after
the merger \citep[see e.g.][]{2020arXiv200106009G}, in which case the
significant old chemical thin disc population found in this work is
unexpected. On the other hand, if a chemical thin disc population were
present in the proto-disc, a significant fraction of these stars
should have been vertically heated and should have large $\zmax$,
similar to thick disc stars. However, the thin disc $\zmax$
distribution does not extend as much as the thick disc to large values
(see Fig.~\ref{fig:apg_old_zmax_kde}, dashed lines). Thus, also the
major merger scenario seems disfavored by our results. Additionally,
possible concerns with this scenario have been recently raised on the
basis of asteroseismic age estimates of thick disc
stars. \cite{2020arXiv200601783M} applied this technique to estimate
ages (within $\sim 10\%$) of metal poor stars ([Fe/H]<-0.5), finding
that part (if not all) of thick disc stars are as old as (or even
slightly older than) stars accreted in the last major merger, thus
concluding that this event was not the main trigger to the thick disc
formation.

Whatever the case may be, these discussions suggest that the thick
disc formation is still puzzling, which should motivate further
studies of alternative scenarios. A compelling possibility is offered
by the clumpy scenario explored in our simulation
(Sec.~\ref{sec:sims}), in which the early galaxy forms clumps
and there are two main star formation channels: $\alpha$-rich stars
are born in clumps and tend to be scattered to high $|z|$ and
reasonably large eccentricities (forming the thick disc), while the
formation of $\alpha$-poor stars is more broadly distributed over the
disc. This scenario has been shown to generate a chemical bi-modality,
geometric thin$+$thick discs and a Splash population in good agreement
with those observed in the MW \citep[][]{2019MNRAS.484.3476C,
  10.1093/mnras/staa065, 2020ApJ...891L..30A}.  Most importantly, an early start for the thin disc
formation, with a large time overlap with the thick disc formation, is a direct prediction of this scenario --
see Fig. 11 in \cite{2019MNRAS.484.3476C}. Additionally, this scenario
also naturally predicts that a fraction of the chemical thick disc
should overlap with the geometric thin disc, i.e. should have low
$\zmax$, as found in this work for the MW -- see
Fig.~\ref{fig:apg_old_zmax_kde}.

 %%%%%%%%%%%%%%%%%%%%%%%%%%%%%%%%%%%%%%%%%%

\subsection{Radial migration: thin disc vs thick disc}

The analysis of Sec.~\ref{sec:time_evolution} showed evidence for the
role of radial migration in the simulation, changing the guiding radii
of stars in nearly-circular orbits typically by
$\langle (\Delta R)^2\rangle^{1/2}\sim 2\kpc$ in a few Gyr. We showed
that a consequence of this is the appearance of a peak in the $\rper$
distribution of old stars, with properties very similar to those found
in the MW. Interestingly, this evidence is present in {\it both} the
chemical thin and thick discs -- see
Fig.~\ref{fig:sim_old_rper_ecc_kde_tsteps}.

Going back to Fig.~\ref{fig:sim_old_GMM_chem_kin_kde}, we can now
evaluate differences in the relevance of radial migration for the thin
and thick discs. As already mentioned, migrators (non-migrators) are
characterized by low (high) eccentricities. On the other hand, inside
the thin or thick discs, the $\zmax$ distribution in our simulation
(Fig.~\ref{fig:sim_old_GMM_chem_kin_kde}-right panels) is similar for
migrators and non-migrators (dashed and dotted curves). This shows
that whether a star will be affected by radial migration depends
essentially on its eccentricity, but not much on its vertical
excursion, in agreement with the conclusions of
\cite{2012MNRAS.422.1363S}. In other words, the extent to which thin
and thick discs are differently affected by radial migration is
determined by their eccentricity distributions, and not by their
different scale heights: the thin disc is more affected by radial
migration because most of its orbits have low eccentricities, while
the fraction of nearly-circular orbits in the thick disc is lower and
it is in this sense that the thick disc is less affected by radial
migration. In fact, for particles in nearly-circular orbits all over
the disc, thin and thick discs are equally affected by radial
migration -- see Fig.~\ref{fig:sim_old_kde_delta_R}. Moreover,
selecting particles at the Solar Neighbourhood, and judging by the
similarity of the $\rper$ KDEs of thin and thick discs at 5 Gyr
(Fig.~\ref{fig:sim_old_rper_ecc_kde_tsteps}) and the similar positions
of the high $\rper$ peak at $13 \Gyr$, one can say that the {\it
  migrators} in both the thin and thick discs are equally influenced
by radial migration, i.e. move by similar amounts in the same time
interval.

Recently, \cite{2020MNRAS.495.3295M} also concluded that the vertical
excursions are not of much relevance to radial migration in simulated
galaxies if the disc is more dominant with respect to the dark matter
halo, while vertical excursions can be more relevant for galaxies with
more dominant halos. Thus, the disc and halo masses relative
contributions can have some importance on our results regarding the
simulated galaxy and the MW. Interestingly, \cite{2017MNRAS.465..798C}
find that baryons dominate the MW's gravitational potential in the
disc out to about the Solar Neighbourhood, which, considering the
conclusions of \cite{2020MNRAS.495.3295M}, reinforces our finding of a
significant migration in the thick disc of the MW.

%%%%%%%%%%%%%%%%%%%%%%%%%%%%%%%%%%%%%%%%%%%%%%%%%%

\subsection{The conundrum of RR~Lyrae in the thin disc}

As discussed in the Introduction, in sequential formation scenarios,
where the thin disc only forms after the thick disc has formed,
RR~Lyrae (which have ages $\gtrsim 10 \Gyr$ ) are not expected to be
found in the thin disc. The recent detection of a population of thin
disc RR~Lyrae then appears as a conundrum in such scenarios
\citep[][]{2020MNRAS.492.3408P}.
One solution of this conundrum was proposed by
\cite{2020arXiv200802280I}.  Using a sample of $\sim \num{70000}$
RR~Lyrae in {\it Gaia}-DR2, they confirmed a sub-population with high
rotational velocity and low $|z|$, typical of the thin disc. They
found that their velocity dispersion is almost a factor of 2 smaller
than that of a sample with age $\tau=10\Gyr$; comparing with the
empirically determined age-velocity dispersion relation of
\cite{2020arXiv200406556S}, they concluded that this population has
velocity dispersion similar to that of a $2\Gyr$-old population. They
suggested that these might be some other population of stars which
mimics RR~Lyrae. We note, however, that a rotationally selected sample
is by definition kinematically biased (towards small velocity
dispersions), while the sample used by \cite{2020arXiv200406556S} to
fit the AVR is complete, and this could potentially lead to the
conclusion of a younger population.

The solution provided by our results is that a thin disc population of
old RR~Lyrae is not a problem in the first place. Our results show
that the existence of these old thin disc stars suggests an early
co-formation of thin and thick discs, a scenario in which thin disc
RR~Lyrae stars are naturally accommodated. In this sense the RR~Lyrae
are no different from other old populations, a fraction of which is
also $\alpha$-poor, thin and rapidly rotating like the thin disc.

%%%%%%%%%%%%%%%%%%%%%%%%%%%%%%%%%%%%%%%%%%%%%%%%%%

\section{Conclusions}
\label{sec:conclusions}
In this work we select, from the \cite{2018MNRAS.481.4093S} catalogue,
old stars ($\tau > 10$ Gyr) within $2\kpc$ from the Sun. After several
quality cuts, our sample consists of $N=\num{23795}$ (turnoffs +
giants) stars with 6-D phase-space coordinates from {\it Gaia}-DR2. We
use these to integrate orbits in an assumed Galactic potential,
determining peri and apo-center radii, eccentricities and maximum
vertical excursions. For a smaller sub-sample of $N=\num{1049}$
(turnoffs + giants) stars, we obtain [Mg/Fe] and [Fe/H] abundances
from APOGEE-DR16 and chemically split this subsample into halo, thin
and thick disc stars. The restricted size of this sample is associated
to fundamental limitations of the observational data currently
available, given the quality cuts required to simultaneously obtain
good age estimates, 6D phase-space information and chemical
abundances. We compare the results with those from a simulated galaxy
which forms clumps in its early evolution. Our conclusions are
summarized below:

\begin{itemize}
\item Chemical thin disc stars represent a significant fraction of our
  sample of old stars, with star counts comparable to those of the
  chemical thick disc -- see Table~\ref{tab:thin_thick_fractions}.  We
  interpret this as evidence that the thin disc formation starts
  early, with a large time overlap with the formation of the thick
  disc.

\item The $\rper$ distribution of old stars has three peaks, one at
  ${\rper\approx 0.59\kpc}$, one at ${\rper\approx 5.4\kpc}$ and
  another at ${\rper\approx 7.14\kpc}$. This confirms the findings of
  \cite{2020MNRAS.492.3408P}, who showed evidence for three $\rper$
  peaks using a sample with 314 RR Lyrae stars.

\item The first peak (at ${\rper\approx 0.59\kpc}$) is associated to
  the stellar halo. The other two peaks have distinct contributions
  from both the thin and thick discs.

\item In both the thin and thick discs, stars contributing to the
  second $\rper$ peak are those with higher eccentricities, while
  stars contributing to the third $\rper$ peak are those on
  nearly-circular orbits. We suggest (and confirm with the simulation)
  that the third peak is produced by stars which migrated from inner
  radii, while the second peak is mostly due to non-migrating stars.

\item In the Solar Neighbouhood, $\sim 1/2$ of the old thin disc stars
  can be classified as migrators, while for the thick disc this
  fraction is reduced to $\sim 1/3$.

\item Thin and thick discs are differently affected by radial
  migration inasmuch as they have different eccentricity
  distributions. Therefore, thick disc stars on nearly-circular orbits
  are as affected by radial migration as thin disc stars in
  nearly-circular orbits.

\item Our results disfavor sequential formation models, such as the
  upside-down scenario, where the thin disc only starts to be
  significant after the thick disc is formed.

\item Among the models which predict an early co-formation of thin and
  thick discs, the clumpy star formation scenario seems to be favoured
  by our results, which is demonstrated by a good agreement between
  observational and simulated trends.
\end{itemize}

%%%%%%%%%%%%%%%%%%%%%%%%%%%%%%%%%%%%%%%%%%%%%%%%%%

\section*{Acknowledgements}
We thank Eugene Vasiliev for help in the use of the {\sc agama}
library. We also thank Giuliano Iorio and Vasily Belokurov by
discussions on their analysis of RR-Lyrae stars, and Stuart Anderson
for highlighting necessary typographic corrections. VPD and LBS are
supported by STFC Consolidated grant \#~ST/R000786/1.  The simulations
in this paper were run at the DiRAC Shared Memory Processing system at
the University of Cambridge, operated by the COSMOS Project at the
Department of Applied Mathematics and Theoretical Physics on behalf of
the STFC DiRAC HPC Facility (www.dirac.ac.uk). This equipment was
funded by BIS National E-infrastructure capital grant ST/J005673/1,
STFC capital grant ST/H008586/1 and STFC DiRAC Operations grant
ST/K00333X/1. DiRAC is part of the National
E-Infrastructure. J.A. acknowledges The World Academy of Sciences and
the Chinese Academy of Sciences for the CAS-TWAS fellowship.
%%%%%%%%%%%%%%%%%%%%%%%%%%%%%%%%%%%%%%%%%%%%%%%%%%

\section*{Data availability}

The Sanders \& Das dataset is publicly available from \href
{https://www.ast.cam.ac.uk/~jls/data/gaia_spectro.hdf5}{this
  URL}. APOGEE-DR16 stellar abundances are available at
\href{https://data.sdss.org/sas/dr16/apogee/spectro/aspcap/r12/l33/allStar-r12-l33.fits}{this
  URL}. The simulation dataset used here is proprietary but can be
shared for limited use on request to V.P.D. (vpdebattista@gmail.com).

%%%%%%%%%%%%%%%%%%%% REFERENCES %%%%%%%%%%%%%%%%%%

% The best way to enter references is to use BibTeX:

\bibliographystyle{mnras}
\bibliography{/Users/lbs/uclan/refs_lbs_uclan}
%\bibliography{refs_lbs_uclan}

\appendix
\section{Silverman test for number of peaks}
\label{sec:silverman}
In this section we explain the test proposed by
\cite{1981JRSSB..43...97S} and used here to test for the number of
peaks in the pericenter radius distribution of our sample. This test
is based on the number of peaks produced by the data KDEs with varying
window widths. A very small window width produces a KDE with many
peaks, possibly over-fitting the data, while a large window width KDE
produces a reduced number of peaks, possibly erasing important
structure in the data. The null hypothesis $H_0$ is that the
distribution has $k$ peaks against the hypothesis $H_1$ that it has
$k+1$ or more peaks. The algorithm is as follows: for each number $k$
of peaks to be tested, starting from one, we define the critical
window width $h_\mathrm{crit}$ as the minimum width to produce at most
$k$ peaks. Then we check the statistical significance of this
$h_\mathrm{crit}$ by bootstrap, re-sampling the data and calculating a
new $h^*_\mathrm{crit}$ for each sample. At a given significance
$\alpha$, we reject $H_0$ if, for an appropriate quantity
$\lambda_\alpha$, the probability
${P(h^*_\mathrm{crit}/h_\mathrm{crit} > \lambda_\alpha) > 1 -
  \alpha}$. The main advantage of this method is that it does not rely
on the Gaussianity assumption inherent in Bayesian Information
Criterion (BIC) and Akaike Information Criterion (AIC) tests under a
Gaussian Mixture Model -- see \cite{2014sdmm.book.....I}. Note that by
``peak'' we mean a local maximum, as opposed to a ``bump'' which is an
interval where the curve is concave (seen from below) but not
necessarily a local maximum.

In \cite{1981JRSSB..43...97S} original test, $\lambda_\alpha = 1$,
which is known to be conservative, i.e. less restrictive in rejecting
models. For $1000$ bootstrap samples, we obtain the probabilities
shown in Table \ref{tab:p_values}, third column. Despite the
conservativeness of the test, models with $k=1$ or $k=2$ peaks are
rejected with high significance. Correcting and calibrating this test,
\cite{Hall_York_2001} obtained $\lambda_\alpha\approx 1.13$ for
$\alpha=0.05$. Although this calibration applies strictly to test for
uni-modality ($k=1$), we also use it to test for larger numbers of
peaks, obtaining the probabilities shown in Table \ref{tab:p_values},
fourth column. Models with $k=1$ or $k=2$ peaks are rejected with high
significance, while a model with $k=3$ peaks cannot be excluded.

\begin{table}
  \begin{center}
    \begin{tabular}{ |c|c|c|c| }
      \hline
      $k$  & $h_\mathrm{crit}$ & $P_\mathrm{Silverman}$ & $P_\mathrm{HY}$ \\
      \hline
      1 & 0.632 & $\approx 1.$ & $\approx 1.$ \\ 
      2 & 0.530 & $\approx 1.$ & $\approx 1.$\\ 
      3 & 0.207 & 0.523 & 0.662 \\
      \hline
    \end{tabular}
    \caption{Results for the test of statistical significance of
      $\rper$ peaks in Fig.~\ref{fig:MW_e_vs_rper_all}. The first
      column ($k$) is the number of peaks to be tested; the second
      column indicates the critical bandwidth, i.e. the minimum to
      produce at most $k$ peaks; the third column shows the
      probability for the number of peaks to be larger than $k$ in the
      original test of \protect\cite{1981JRSSB..43...97S}; the forth
      column shows the probabilities after the correction/calibration
      of \protect\cite{Hall_York_2001}. Models with $P>0.95$ can be
      excluded.}
        \label{tab:p_values}
  \end{center}
\end{table}

\section{Geometric cuts and selection effects}
\label{sec:cuts}

In Sec.~\ref{sec:halo_thin_thick}, we present the thin and thick disc
fractions in our sample of old stars in the fiducial cut (ages
$\tau>10\Gyr$ and distances $d<2\kpc$). This cut obviously introduces
selection effects and those fractions should not be taken at face
value. In this appendix, we make a simple evaluation of the selection
effects by varying the cuts and recalculating those
fractions. Additionally to the fiducial cut, we select stars with
$\tau > 11\Gyr$ and within the Solar cylinder, i.e.
($|R\!-\!R_\odot| < 2\kpc$) and different cuts in
$|z|$. Table~\ref{tab:thin_thick_fractions} shows the number counts in
the cross-match with APOGEE and the thin and thick disc fractions for
different combinations of these cuts. In general, thin disc stars
correspond to $\sim 20 - 30\%$ of the samples (with slightly larger or
smaller fractions for cuts intended to suppress or increase the thin
disc contribution), while the thick disc corresponds to
$\sim 40 - 55\%$. Regardless of exact values, this indicates that the
thin disc represents a significant component of the old Milky Way.

\begin{figure}
  \includegraphics[width=\columnwidth]{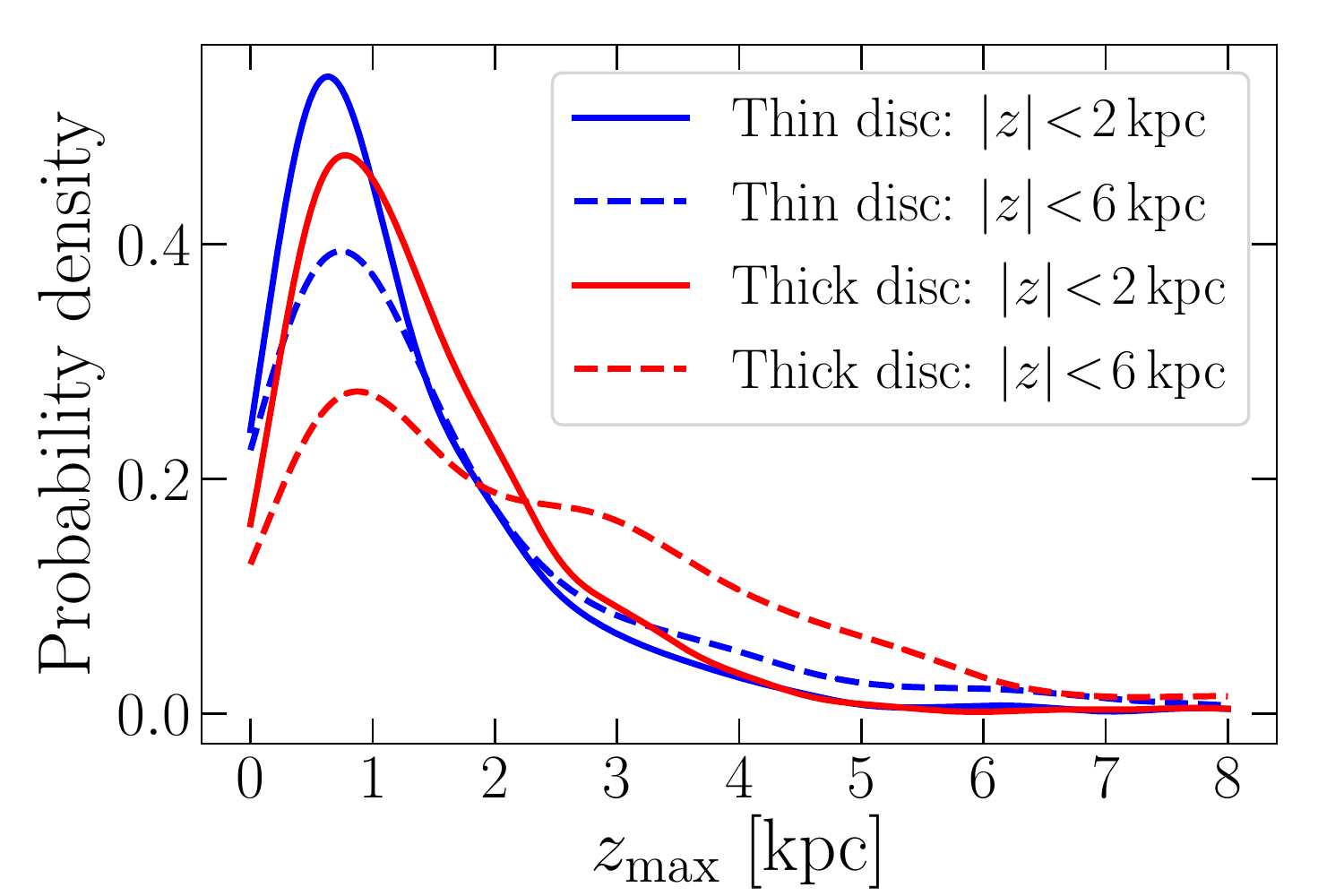}
  \caption{KDE of $\zmax$ for old stars in the thin and thick discs
    (blue and red, respectively) in the cross-match with APOGEE-DR16
    within cylindrical regions around the Sun (${R-R_\odot < 2\kpc}$)
    -- see Table~\ref{tab:thin_thick_fractions} for details. While the
    thin disc KDE is changed for the different cuts, the thick disc
    extends to much larger $\zmax$ when we select stars with
    $|z|<6\kpc$. However, even in this case, a significant fraction of
    chemical thick disc stars are confined to the geometric thin disc,
    i.e. have low $\zmax$.}
  \label{fig:apg_old_zmax_kde}
\end{figure}

Another finding of this paper is the significant fraction of chemical
thick disc stars in the geometric thin disc region, i.e. at small
$\zmax$. This could in principle be flawed by the suppression of thick
disc stars with large $\zmax$ in the fiducial
cut. Fig.~\ref{fig:apg_old_zmax_kde} shows that the thick disc indeed
extends to high $\zmax$ when selecting stars with $|z|<6\kpc$ (dashed
red curve). However, its distribution still peaks at
$\zmax \sim 1\kpc$ and has a significant low $\zmax$ population.

% DR16:

\begin{table}
  \begin{center}
    \begin{tabular}{ |c|c|c|c|c|c|c| }
      \toprule
      $\tau$ & $d$  & $|R\!-\!R_\odot|$& $|z|$ & $N$ & Thick & Thin \\
      $\mathrm{[Gyr]}$ & $\mathrm{[kpc]}$ & $\mathrm{[kpc]}$
                                       &$\mathrm{[kpc]}$ & & & \\
      \hline
      \multirow{6}{*}{$\!>\!10$} & $\!<\!0.5$ & --&-- & 222 & $40\%$ & $50\%$ \\ %\cline{2-6}
             & ${\bf \!<\! 2}$ & {\bf --} & {\bf --} & {\bf 1049} & ${\bf 54\%}$ & ${\bf 29\%}$ \\\cline{2-7}
             & \multirow{4}{*}{--} & \multirow{4}{*}{$<2$} & $<2$ & 1489 & $52\%$ & $26\%$ \\%\cline{3-6}
             & & & $<6$ & 2238 & $46\%$ & $17\%$ \\%\cline{3-6}
             & & &\textcolor{red}{$0.3\!<\!|z|\!<\!6$} & \textcolor{red}{1909} & \textcolor{red}{$48\%$} & \textcolor{red}{$12\%$} \\%\cline{3-6}
             & & & \textcolor{blue}{$<0.3$} & \textcolor{blue}{330} & \textcolor{blue}{$37\%$} & \textcolor{blue}{$54\%$} \\
      \midrule
      \multirow{5}{*}{$\!>\!11$} & $\!<\!0.5$ & --& -- & 124 & $46\%$ & $37\%$ \\ %\cline{2-6}
             & ${\bf \!<\!2}$ &{\bf --} & {\bf --} & {\bf 573} & ${\bf 54\%}$ & ${\bf 25\%}$ \\ \cline{2-7} 
             & \multirow{4}{*}{--} & \multirow{4}{*}{$<2$} & $<2$ & 759 & $48\%$ & $25\%$
      \\ %\cline{3-6}
             & & & $<6$ & 1036 & $41\%$ & $18\%$ \\ %\cline{3-6}
             & & & \textcolor{red}{$0.3\!<\!|z|\!<\!6$} & \textcolor{red}{849} & \textcolor{red}{$41\%$} & \textcolor{red}{$12\%$} \\ %\cline{3-6}
             & & & \textcolor{blue}{$<0.3$} & \textcolor{blue}{187} & \textcolor{blue}{$38\%$} & \textcolor{blue}{$48\%$} \\
      \bottomrule
    \end{tabular}
    \caption{Simple estimates of selection effects in the cross-match
      with APOGEE-DR16: for different age and geometric cuts, the
      right-hand columns show the fractions of thin and thick disc
      stars compared to the total sample (including halo stars, not
      shown). The fiducial cut for $\tau > 10\,\mathrm{Gyr}$, and the
      corresponding one with $\tau > 11\,\mathrm{Gyr}$, are shown in
      boldface. Neglecting cuts explicitly intended to suppress the
      thin disc (red), this component contribution is $\gtrsim 20\%$
      of the star counts.}
        \label{tab:thin_thick_fractions}
  \end{center}
\end{table}

%%%%%%%%%%%%%%%%%%%%%%%%%%%%%%%%%%%%%%%%%%%%%%%%%%

%%%%%%%%%%%%%%%%% APPENDICES %%%%%%%%%%%%%%%%%%%%%

%%%%%%%%%%%%%%%%%%%%%%%%%%%%%%%%%%%%%%%%%%%%%%%%%%

% Don't change these lines
\bsp	% typesetting comment
\label{lastpage}
\end{document}